\documentclass[fleqn,usenatbib]{mnras}

\usepackage{newtxtext,newtxmath}
\usepackage[T1]{fontenc}
\usepackage{ae,aecompl}
\usepackage{xcolor}

\usepackage{graphicx}	% Including figure files
\usepackage{amsmath}	% Advanced maths commands
\title[Spectroscopic IR reverberation: Mrk~509]{The first spectroscopic IR reverberation programme on Mrk~509}

\author[J. A. J. Mitchell et al.]{J. A. J. Mitchell$^1$\thanks{E-mail: jake.a.mitchell@durham.ac.uk}\thanks{Visiting Astronomer at the Infrared Telescope Facility, which is operated by the University of Hawaii under contract NNH14CK55B with the National Aeronautics and Space Administration.}, M. J. Ward$^1$\footnotemark[2], D. Kynoch$^2$\footnotemark[2], J. V. Hern\'andez Santisteban$^3$,
\newauthor
K. Horne$^3$, J.-U. Pott$^4$, J. Esser$^4$, P. Mercatoris$^4$, C. Packham$^{5,6}$, G. J. Ferland$^7$, A. Lawrence$^8$, 
\newauthor
T. Fischer$^{9}$, A. J. Barth$^{10}$, C. Villforth$^{11}$, H. Winkler$^{12}$ \\ 
$^1$Centre for Extragalactic Astronomy, Department of Physics, Durham University, South Road, Durham, DH1 3LE, UK \\ 
$^{2}$Astronomical Institute, Academy of Sciences, Bo\v{c}n\'{i} II 1401, 14131 Prague, Czech Republic \\
$^3$SUPA Physics and Astronomy, University of St. Andrews, Fife, KY16 9SS, UK \\
$^4$Max Planck Institut f\"ur Astronomie, K\"onigstuhl 17, D-69117 Heidelberg, Germany \\
$^5$Department of Physics \& Astronomy, University of Texas at San Antonio, One UTSA Circle, San Antonio, TX 78249, USA \\
$^6$National Astronomical Observatory of Japan, 2-21-1 Osawa, Mitaka, Tokyo 181-8588, Japan \\
$^7$Department of Physics and Astronomy, University of Kentucky, Lexington, KY 40506, USA \\
$^8$Institute for Astronomy, University of Edinburgh, Royal Observatory, Blackford Hill, Edinburgh, EH9 3HJ, UK \\
$^9$Space Telescope Science Institute, 3700 San Martin Drive, Baltimore, MD 21218, USA \\
$^{10}$Department of Physics and Astronomy, 4129 Frederick Reines Hall, University of California, Irvine, CA, 92697-4575, USA \\
$^{11}$Department of Physics, University of Bath, Claverton Down, Bath BA2 7AY, UK \\
$^{12}$Department of Physics, University of Johannesburg, PO Box 524, 2006 Auckland Park, South Africa}

\date{Accepted 2024 March 8. Received 2024 February 9; in original form 2022 October 1}

\pubyear{2022}

\begin{document}

\def\la{\mathrel{\hbox{\rlap{\hbox{\lower4pt\hbox{$\sim$}}}\hbox{$<$}}}}
\def\ga{\mathrel{\hbox{\rlap{\hbox{\lower4pt\hbox{$\sim$}}}\hbox{$>$}}}}

%emission line definitions
\font\sevenrm=cmr7
\def\OIII{[O~{\sevenrm III}]}
\def\FeII{Fe~{\sevenrm II}}
\def\FeIIf{[Fe~{\sevenrm II}]}
\def\SIII{[S~{\sevenrm III}]}
\def\HeI{He~{\sevenrm I}}
\def\HeII{He~{\sevenrm II}}
\def\NeV{[Ne~{\sevenrm V}]}
\def\OIV{[O~{\sevenrm IV}]}

\def\iraf{{\sevenrm IRAF}}
\def\mpfit{{\sevenrm MPFIT}}
\def\galfit{{\sevenrm GALFIT}}
\def\mapspec{{\sevenrm mapspec}}
\def\cream{{\sevenrm CREAM}}
\def\javelin{{\sevenrm JAVELIN}}
\def\clumpy{{\sevenrm CLUMPY}}
\def\cloudy{{\sevenrm CLOUDY}}
\def\astroimagej{{\sevenrm AstroImageJ}}
\def\banzai{{\sevenrm BANZAI}}
\def\orac{{\sevenrm ORAC}}
\def\python{{\sevenrm python}}
\def\prepspec{{\sevenrm prepspec}}
\def\GW{{\sevenrm GW92}}
\def\spextool{{\sevenrm Spextool}}

\label{firstpage}
\pagerange{\pageref{firstpage}--\pageref{lastpage}}

\maketitle

\begin{abstract}

Near IR spectroscopic reverberation of Active Galactic Nuclei (AGN) potentially 
allows the IR broad line region (BLR) 
to be reverberated alongside the disc and dust continua, while the spectra can also reveal details of dust astro-chemistry. Here we describe results of a short pilot study (17 near-IR spectra over a 183 day period) for Mrk 509. The spectra give a luminosity-weighted dust radius of \mbox{$\langle R_{\rm d,lum} \rangle = 186\pm4$~light-days} for blackbody (large grain dust), consistent with previous (photometric) reverberation campaigns, whereas carbon and silicate dust give much larger radii. 
We develop a method of calibrating spectral data in objects where the narrow lines are extended beyond the slit width. We 
demonstrate this by showing our resultant
photometric band lightcurves are consistent with previous results, with a 
hot dust lag at >40 days in the $K$-band, clearly different from the accretion disc response at <20 days in the $z$-band. We place this limit of 40 days by demonstrating clearly that the modest variability that we do detect in the $H$ and $K$-band does not reverberate on timescales of less than 40 days. 
We also extract the Pa$\beta$
line lightcurve, and find a lag which is consistent with the optical BLR H$\beta$ line of $\sim70-90$ days. This is important as direct imaging of the near-IR BLR is now possible in a few objects, so we need to understand its relation to the better studied optical BLR.

\end{abstract}

\begin{keywords}
galaxies: Seyfert -- infrared: galaxies -- quasars: emission lines -- quasars: individual: Mrk~509 
\end{keywords}

\section{Introduction}

Active Galactic Nuclei (AGN) are powered by mass accretion onto a central Supermassive Black Hole (SMBH), and are the most luminous compact objects in the Universe. They exhibit a wide range of emission characteristics resulting in an AGN zoo of differing classifications. One of the fundamental distinctions between categories is that which separates AGN with both narrow and broad emission lines seen in their optical spectra (Type 1) from those which only display narrow lines (Type 2). Historically unified schemes proposed a basic AGN anatomy consisting of a central SMBH, a thin accretion disc, separate compact broad and extended narrow line regions and an outer dusty toroidal structure, with each of these components existing on different physical size scales. It has been argued, through these unified schemes, that orientation effects resulting in the obscuration of the Broad Line Region (BLR) for certain viewing angles is the physical mechanism separating AGN into type 1 and type 2 sources \citep{Law87, Ant93,Urry95, Netzer15}. Originally the dichotomy between type 1 and 2 AGN was solely explained by these orientation effects. However the predominant view now is more holistic, with differences in emission characteristics attributed to numerous factors, such as mass accretion rate and black hole mass in addition to orientation \citep{Jin12,KubotaDone2018}.

Despite this slight paradigm shift, much evidence for the presence of a large dusty obscuring structure, still commonly referred to as the torus for ease, remains. Strong infrared continuum emission at $\mathrm{\lambda}$ > 1 $\mathrm{\mu}$m, thought to be reprocessed optical/UV emission from the accretion disc, and observations of broad emission lines in the polarised spectra of some type-2 AGN \citep{ant85,Ant93,Coh99,Lum01, Tran03} are prime examples of evidence for the torus. For a small subset of AGN, constraints on the spatial extent of the torus have been placed at (0.1-6) pc through near-IR, mid-IR \citep{Tri09,Pott10,Kish11,Bur13, Kish13} and sub-mm interferometric observations \citep{Garcia16,Imanishi18} along with high angular resolution mid-IR imaging \citep{Pack05,Ramos11}. These findings are roughly consistent with a power-law size-luminosity relation $r_{\mathrm{dust}}$ = $L^{\mathrm{0.5}}$ \citep{Kish09,Kish11}. Modelling the torus has also provided a useful probe of this structure. Through the fitting of infrared SED's, using the \clumpy{} torus model \citep{AH2011}, and spectroscopy, tight constraints have been placed on a range of torus properties. This and other studies indicate that the torus is of a clumpy geometry \citep{RA11,Ichikawa15}. However for the general case the precise structure and geometry of the torus remains unknown, largely due to the size scales in question, rendering direct imaging impossible with current technology for the vast majority of sources.

Where interferomtery is not possible, the inner dusty structures of AGN can be probed with reverberation mapping techniques, which are predominantly used in mapping the spatial extent of the broad line region in AGN, but can also be used to measure the physical size of other components such as the accretion disc or hot dust \citep{Clavel89,Glass92,Sit93,Nel96,Okn01,Glass04}. This method assumes that the variable near-IR continuum arises from hot dust reprocessing of the variable optical/UV emission from the accretion disc and therefore the inner edge of the torus can be measured as the time lag between these two variable signals.  This method is well tested, with ~20 AGNs having been reverberation mapped in the near-IR through long term coordinated optical and near-IR monitoring campaigns \citep{Clavel89,Sit93,Nel96,Okn01,Glass04,Min04,Sug06,Kosh14,Vazquez15}. Dust radii measured in this way are systematically smaller than radii found with near-IR interferometry \citep{Okn01,Kish07,Nen08a}. The same discrepancy is seen between reverberated radii and those determined by the dust sublimation temperature and AGN bolometric luminosity. Previous studies have attributed this offset to the dust geometry, advocating for a bowl-shaped rather than a spherical (doughnut-like) structure, caused by the anisotropy of the accretion disc emission \citep{Kaw10,Kaw11}. This shape would also be seen if the dust formed part of a radiatively accelerated outflow from the disc, or a disc wind \citep{Hon17}.

Reverberation mapping has most commonly been carried out with photometric data, largely due to the relative ease in obtaining high cadence lightcurves for multiple objects simultaneously. However photometric reverberation mapping has limitations that can be overcome by using spectroscopy instead. The lightcurves are not true reflections of the continuum flux variability as, due to fairly wide bandwidth filters, they are often contaminated by broad line emission, which superimpose their own intrinsic variability originating from the BLR, and show reverberation, but from different radii than the hot dust emission \citep{Pet04,Kosh14}. With time-resolved spectroscopic data these line profiles can be reliably separated from the continuum emission, providing not only an uncontaminated continuum lightcurve, but also allowing for the assembly of simultaneous emission-line lightcurves. This allows us to investigate the connection between variations of emission-line flux and profile shape simultaneously with dust continuum variations. In the scheme that the BLR is limited by the dusty torus, we could even gain insights into the shape of the BLR, current evidence would suggest that the BLR, like the torus could be bowl shaped \citep{Goad12}. Most BLR reverberation mapping is carried out on the optical lines, however, and so is most probably sampling a smaller size scale than the most extended part of the BLR.  In this work we extend emission-line reverberation studies to infrared lines, namely Pa$\mathrm{\epsilon}$ and Pa$\mathrm{\beta}$. In addition to line measurements, spectroscopy also makes the simultaneous measurement of dust temperature possible, making astro-chemical analysis possible through sublimation arguments. This technique is now viable due to the advent of short-cross dispersed near-IR spectrographs mounted on medium sized telescopes, and has been utilised in \citet{L19} which focused on NGC 5548. 

In this work we present our findings from a 7-month long spectroscopic near-IR reverberation campaign on the nearby, well studied AGN Mrk 509 conducted between 2019 May to November.  In this paper, we focus mainly on the near-IR continuum emission, but also investigate the variability of Pa$\mathrm{\beta}$ and Pa$\mathrm{\epsilon}$. We ultimately are able to place lower limits on the hot dust and Paschen lags and thereby demonstrate the feasibility of measuring both of these values with a better sampled data set.

We adopt here cosmological parameters $H_0 = 70$~km~s$^{-1}$~Mpc$^{-1}$, $\Omega_{\rm M}=0.3$, and $\Omega_{\Lambda}=0.7$, which give a luminosity distance to Mrk 509 of 150 Mpc and an angular scale at the source of 682 pc per arcsec.

\section{The science target}

This study focuses on the Type-1 AGN Mrk 509. We selected this target for several reasons: (i) it has a previously measured, short and variable dust response time of ($\mathrm{126\pm11}$) days \citep{Kosh14}, therefore within our 6 month baseline of observations we should be able to recover a hot dust lag measurement; (ii) it has clearly discernible broad emission lines in the Paschen series allowing us to separate narrow- and broad-line components; (iii) multiple reverberation campaigns have determined its black hole mass. \citet{Pet04} reported the black hole mass to be (1.43$\mathrm{\pm}$0.12)$\mathrm{\times 10^{8} \ M_{\odot}}$. In addition Mrk 509 has a spatially resolved torus radius of ($\mathrm{296\pm31}$) lt-days from near-IR interferometry \citep{grav20}.

Mrk 509 (J2000 sky coordinates \mbox{R.A. $20^h 44^m 09.8^s$} and \mbox{Decl. $-10^\circ 43' 24.7''$}) can be observed from both the northern and southern hemispheres. Its relatively low redshift of $z=0.0344$, enables us to sample the hot dust component and several near-IR coronal lines using a cross-dispersed near-IR spectrum. This AGN inhabits a compact host galaxy, perhaps of type S0 \citep{kriss11} and is relatively bright in the near-IR (2MASS $J = 11.6$~mag, $K = 10.0$~mag; \citet{2MASS}) ensuring good quality near-IR spectra are possible using reasonable exposure times on a 4~m class telescope. 

\section{The observations}

Our aim is to measure a reverberation signal in the near-IR hot dust emission and the Paschen emission lines of Mrk 509, and to measure the dust temperature. We therefore monitor Mrk 509 using near-IR spectroscopy and also require well sampled optical photometry to trace the accretion disc variations.   

\subsection{The near-IR spectroscopy} \label{spectroscopy}

Using the Spex spectrograph at the 3~m NASA Infrared Telescope Facility (IRTF) on Manaukea, Hawaii,  we observed the source Mrk 509 between 2019 May and November (semesters 2019A and 2019B) \citep{spex}. Having requested 36 nights of observing to obtain a cadence of $\sim$1 week, we were scheduled 24, 2-2.5 hour, observing windows between 2019 April and November (resulting in successful observations between May and November 2019).  We obtained a total of 17 near-IR spectra as we lost seven of the scheduled observations due to weather and site inaccessibility. The journal of observations is listed in Table \ref{obslog}.

All observations were carried out using the (0.7-2.55) $\mu$m short cross-dispersed mode (SXD) equipped with the 0.3"x15" slit, oriented at the parallactic angle. Using this narrow slit allowed us to minimise the contamination from the host galaxy, whilst giving a sufficiently high spectral resolution, of $R$=2000 or full width at half maximum (FWHM) $\sim$ 150 km s$^{-1}$, to discern clearly any narrow and broad emission lines and study their profiles. We observed the source using a standard ABBA nodding pattern. As the source is not extended we nodded along the slit rather than on sky, therefore obtaining an image of the target in both the A and B frames from which the background subtraction could be carried out. The usual on-source exposure time was 32$\mathrm{\times}$120s. On seven of the seventeen observing runs we obtained fewer exposures than this due to either weather or technical issues. We also note that on 2019 September 10 we achieved 38 exposures due to spare time. This exposure time was chosen in order to guarantee high enough (S/N) to reliably measure not only the broad, but also the narrow emission line profiles, which can be key in performing photometric corrections to the spectra, as discussed in Section~\ref{rescaling}. 

We observed a standard star to correct for telluric absorption and for flux calibration. Therefore after the science target we observed a nearby (in airmass and position) A0 V star, HD 203893 which has accurate optical magnitudes (\textit{B} = 6.89 and \textit{V} = 6.84). This spectral type has a well defined spectral shape into the infrared and using the quoted optical magnitudes along with the \spextool{} reduction pipeline this shape was used to perform the IR telluric correction. Where possible we took flats and arcs after the science target. Photometric corrections to the spectra in order to investigate intrinsic variability is carried out through a separate process detailed in Section~\ref{rescaling}.

\begin{figure*} 
\centerline{
\includegraphics[scale=0.6, angle=-90, clip=true, bb=40 30 510 770]{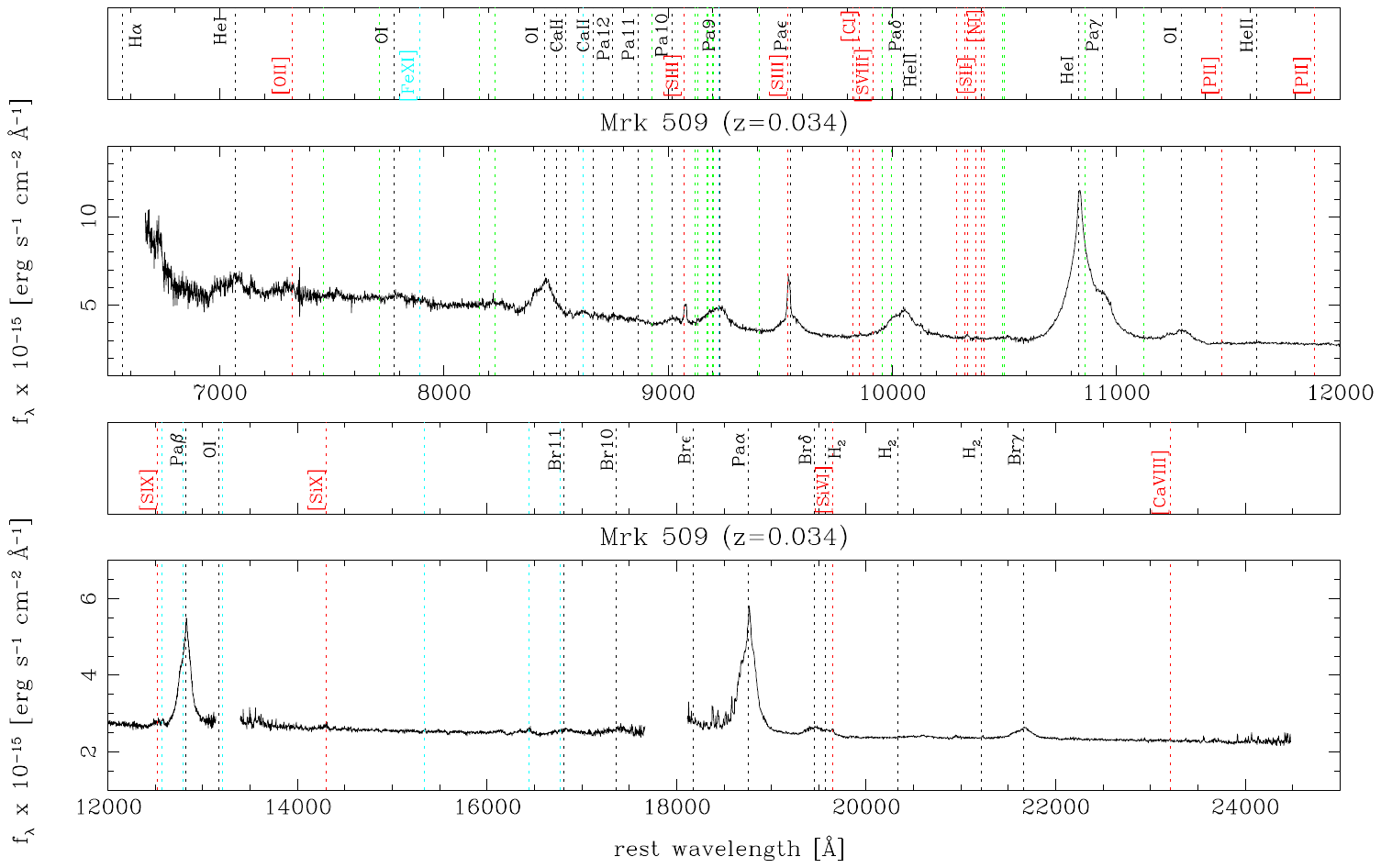}
}
\caption{\label{irtfspec} IRTF SpeX near-IR spectrum from 2019 May 9 shown as observed flux versus rest-frame wavelength. Emission lines listed in Table 4 of \citet{L08a} are marked by dotted lines and labeled; black: permitted transitions, green: permitted Fe~{\sc ii}~multiplets (not labeled), red: forbidden transitions and cyan: forbidden transitions of iron (those of $[$Fe~{\sc ii}$]$~not labeled).}
\end{figure*}

The data were reduced using \spextool{} (version 4.1) a software package created for use by Spex users and based in Interactive Data Language (IDL) \citep{spextool}. This package was used on both the science target and telluric standard. Spextool carries out all of the procedures required in the production of fully reduced spectra. This includes the preparation of calibration frames, the subtraction of flat fields, the fitting and subtraction of local background, interactive spectral extraction from science frames using an optimally weighted extraction scheme \citep{Horne86a}, telluric correction of the science spectra, merging of orders, and combination of the different frames into a single reduced science spectrum for each observing window. The final spectrum was corrected for Galactic extinction using a value of $A_{\rm V}=0.079$, which we derived from the Galactic hydrogen column densities published by \citet{DL90}. In Figure~\ref{irtfspec}, we show the spectrum from 2019 May 9 as a representative example.

\begin{table*}
\caption{\label{obslog} 
IRTF Journal of observations}
\begin{tabular}{lrccrrrrccclc}
\hline
Observation &MJD & exposure & airmass & aperture & PA & \multicolumn{3}{c}{continuum S/N}  & Standard Star &seeing  &cloud \\
Date & &(s)  && (arcsec$^2$) & ($^{\circ}$) & $J$ & $H$ & $K$  & airmass  & (arcsec) & condition\\
(1) & (2) & (3) & (4) & (5) & (6) & (7) & (8)  & (9) & (10) & (11) & (12) \\
\hline
2019 May 9  & 58613 & 32$\times$120 & 1.39  & 0.3$\times$1.3 & 293 & 40 & 92 & 103  & 1.32 & 0.8 & clear    \\
2019 May 20 & 58624 & 32$\times$120 & 1.24  & 0.3$\times$1.4 & 106 & 48 & 82 & 117  & 1.23 & 0.8 & cirrus   \\
2019 May 30 & 58634  & 32$\times$120 & 1.18  & 0.3$\times$1.3 & 100 & 52 & 93 & 117  & 1.21 & 0.5 & photom.  \\
2019 Jun 8  & 58643  & 32$\times$120 & 1.16  & 0.3$\times$1.3 &  94 & 45 & 81 & 111  & 1.21 & 0.5 & photom.  \\
2019 Jun 14 & 58649  & 32$\times$120 & 1.16  & 0.3$\times$1.2 & 266 & 40 & 73 & 109  & 1.24 & 0.5 & photom.  \\
2019 Jun 18 & 58653  & 32$\times$120 & 1.17  & 0.3$\times$1.2 &  83 & 49 & 85 & 114  & 1.26 & 0.7 & clear    \\
2019 Jul 15 & 58679  & 32$\times$120 & 1.26  & 0.3$\times$1.2 & 108 & 46 & 85  & 103  & 1.24 & 0.4 & photom.  \\
2019 Aug 26 & 58721 & 24$\times$120 & 1.39  & 0.3$\times$1.1 & 113 & 52 & 84 & 106  & 1.38 & 0.5 & cirrus   \\
2019 Sep 2  & 58728 & 16$\times$120 & 1.22  & 0.3$\times$1.5 & 104 & 35 & 75 &  93  & 1.27 & 0.6 & cirrus   \\
2019 Sep 10 & 58736 & 38$\times$120 & 1.17  & 0.3$\times$1.2 &  97 & 56 & 84 & 110  & 1.21 & 0.7 & cirrus   \\
2019 Sep 19 & 58745 & 24$\times$120 & 1.16  & 0.3$\times$1.1 & 100 & 54 & 75  &  95  & 1.21 & 0.5 & cirrus.  \\
2019 Sep 26 & 58752 & 24$\times$120 & 1.31  & 0.3$\times$1.4 & 230 & 56 & 77 & 115  & 1.31 & 0.7 & cirrus   \\
2019 Oct 1  & 58757 & 26$\times$120 & 1.30  & 0.3$\times$1.3 & 109 & 59 & 75 & 103  & 1.27 & 0.6 & clear    \\
2019 Oct 21 & 58777 & 32$\times$120 & 1.17  & 0.3$\times$1.6 & 265 & 51 & 59 &  97  & 1.25 & 1.2 & cloudy   \\
2019 Oct 31 & 58787 & 30$\times$120 & 1.19  & 0.3$\times$1.4 &  79 & 44 & 76 & 119   & 1.32 & 0.8 & clear   \\
2019 Nov 4  & 58791 & 16$\times$120 & 1.21  & 0.3$\times$1.2 &  76 & 40 & 67 & 109  & 1.27 & 0.7 & clear   \\
2019 Nov 14 & 58801 & 24$\times$120 & 1.23  & 0.3$\times$1.4 & 255 & 37 & 61 &  95  & 1.12 & 0.5 & clear   \\
\hline
\end{tabular} 

\parbox[]{18cm}{The columns are: (1) Universal Time (UT) date of observation; (2) Modified Julian Date (MJD); (3) exposure time; (4) mean airmass; (5) extraction aperture; (6) slit position angle, where PA$=0^{\circ}$ corresponds to east-west orientation and is defined east through north; S/N in the continuum over $\sim 100 \mathrm{\Angstrom}$ measured at the central wavelength of the (7) \textit{J}, (8) \textit{H}, and (9) \textit{K}-bands; for the telluric standard star; (10) mean airmass; (11) seeing in the $K$ band; (12) cloud condition. All observations were taken contemporaneously with the telluric standard star HD 203893.}

\end{table*}

\subsection{Complementary Optical photometry}

\begin{figure*} 
\centerline{
\includegraphics[scale=0.26, clip=true]{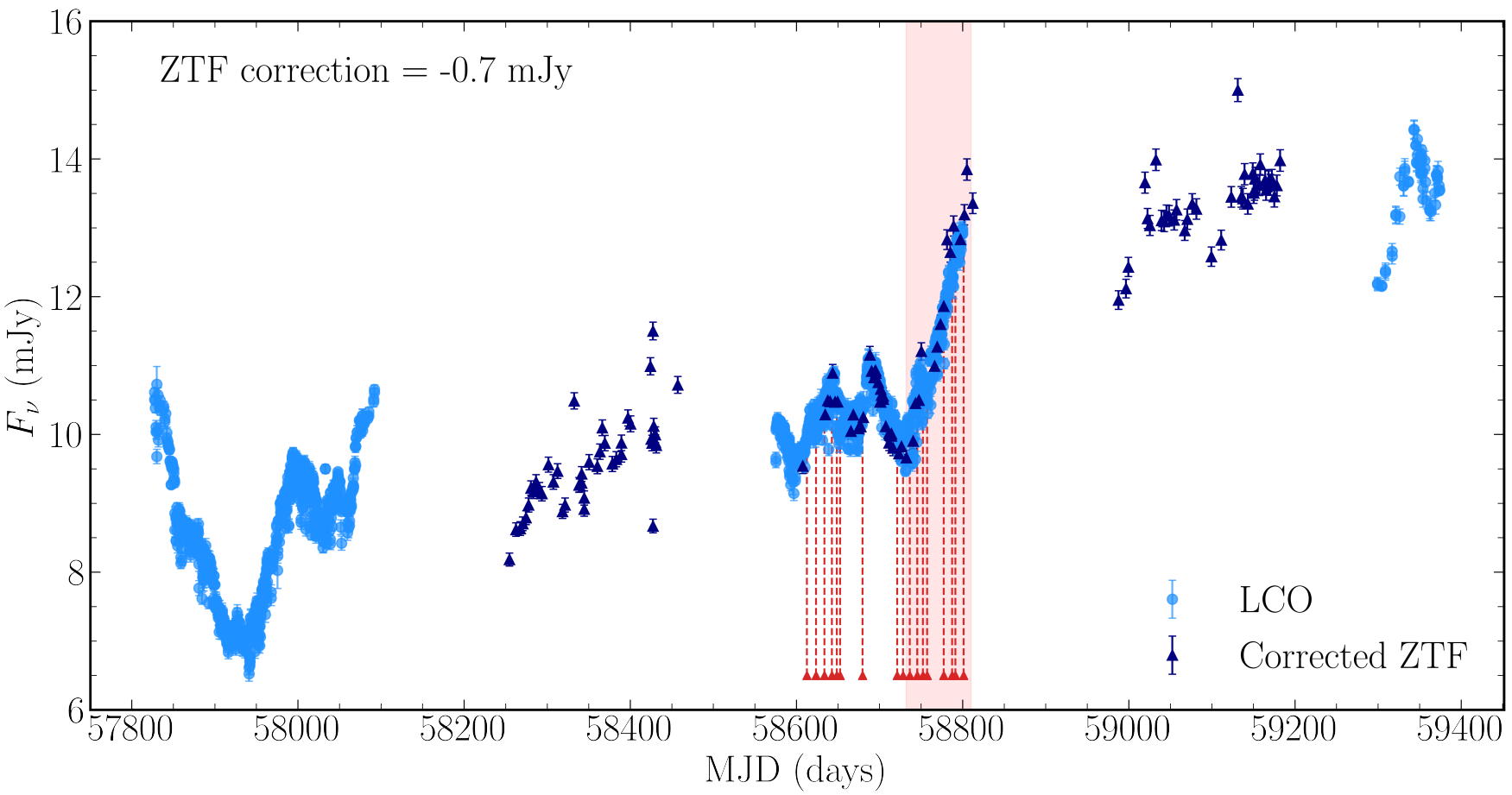}
}
\caption{\label{ztf_lco} LCO $g$-band observations of Mrk 509 with complementary corrected ZTF $g$ band data. The most significant variability feature measured contemporaneously with our near-IR spectra is shaded in red. Archival LCOGT data (MJD<58200) is shown to extend the time baseline. Spectral epochs are marked by red arrows. }
\end{figure*}

\subsubsection{The LCOGT observations}

Between 2019 April 2 and November 12 we monitored Mrk~509 in the optical with the 1~m and 2~m robotic telescopes of the Las Cumbres Observatory \citep[LCOGT;][]{LCOGT} almost daily in four bands ($g$, $r$, $i$ and $z_s$). This baseline of observations begins $\sim$1 month before our near-IR spectra and ends 2 days before the end of our near-IR campaign. We use here mainly the lightcurves in the $g$ and $z_s$ filters, which have a central wavelength of 4770~\AA~and 8700~\AA~and a width of 1500~\AA~and 1040~\AA~respectively. The Sinistro cameras on the 1~m telescopes have a field-of-view (FOV) of $26.5' \times 26.5'$ and a plate scale of $0.389''$ per pixel. The Spectral cameras on the 2~m telescopes have a FOV of $10.5' \times 10.5'$ and a plate scale of $0.304''$ per pixel. 

The frames were first processed by LCOGT's \banzai~pipeline \citep{banzai} in the usual way (bias and dark subtraction, flat-fielding and image correction) and were subsequently analysed with our custom-made pipeline, which has been described in detail by \citet{Hernandez20}. In short, after performing aperture photometry with a diameter of $7''$ and subtracting a background model, we produced stable light curves by constructing a curve of growth using the standard stars on each individual frame and measuring the correction factors required to bring all different light curves to a common flux level. This approach mimicked extraction from an azimuthally averaged point-spread function (PSF), but without actually performing a PSF extraction. A colour correction and correction for atmospheric extinction were applied before the photometric calibration. Finally, we used comparison stars in each field to perform an image zero-point calibration at each epoch. In Figure~\ref{ztf_lco}, we show the $g$ and in Figure~\ref{z_check} the $z_s$ light-curve, with the latter overlapping in wavelength with the near-IR spectrum.

\subsubsection{The ZTF data}

The LCO $g$-band lightcurve has been supplemented with data taken from the Zwicky Transient Facility (ZTF) \citep{ZTF}. These data were corrected to match the flux scale of the LCO observations by interpolating the LCO flux lightcurve to the ZTF observation times, calculating a mean flux difference, and then subtracting this from the ZTF lightcurves. This correction value was $\sim$0.7 mJy. The resulting ZTF lightcurve is displayed in Figure \ref{ztf_lco} alongside the LCO observations. The ZTF data fills large gaps present in the LCO data-set, and therefore provides a helpful constraint on the behaviour of Mrk 509 during these time periods.   

\subsubsection{The GROND observations}

\begin{table*}
\caption{\label{grondphot} 
Observed optical and near-IR nuclear fluxes from GROND photometry}
\begin{tabular}{lcccccc}
\hline
Observation & $g'$ & $r'$ & $i'$ & $J$ & $H$ & $K$ \\
Date & $\lambda_{\rm eff}=4587$~\AA & $\lambda_{\rm eff}=6220$~\AA & $\lambda_{\rm eff}=7641$~\AA & $\lambda_{\rm eff}=12399$~\AA & $\lambda_{\rm eff}=16468$~\AA & $\lambda_{\rm eff}=21706$~\AA \\ 
& (mJy) & (mJy) & (mJy) & (mJy) & (mJy) & (mJy) \\
\hline
2019 Aug 26 & 9.7 & 13.3 & 10.9 & 17.4 & 22.7 & 40.3 \\
\hline  
\end{tabular} 
\end{table*}

We obtained simultaneous optical and near-IR photometry in seven bands with the Gamma-Ray Burst Optical and Near-Infrared Detector \citep[GROND;][]{grond} mounted on the MPG~2.2~m~telescope, La Silla, Chile. The observations were taken on 2019 August 26, simultaneous with one of our near-IR spectra. The optical channels have a FOV of $5.4' \times 5.4'$ and a plate scale of $0.158''$ per pixel, whereas both the FOV and plate scale of the near-IR channels is larger ($10' \times 10'$ and $0.6''$ per pixel, respectively). The filter passbands are relatively wide ($\sim 1000 - 3000$~\AA). Images were reduced with the \iraf-based pipelines developed by R. Decarli and G. De Rosa \citep{Morg12}, modified and extended for our purpose. After bias and dark subtraction and flat-fielding of each individual image, a scaled median sky image constructed from separate sky exposures was subtracted from each frame. The dither step size was $18''$ for both the science and sky images, with five dither positions obtained for each. 

\begin{figure}
\centerline{
\includegraphics[scale=0.31]{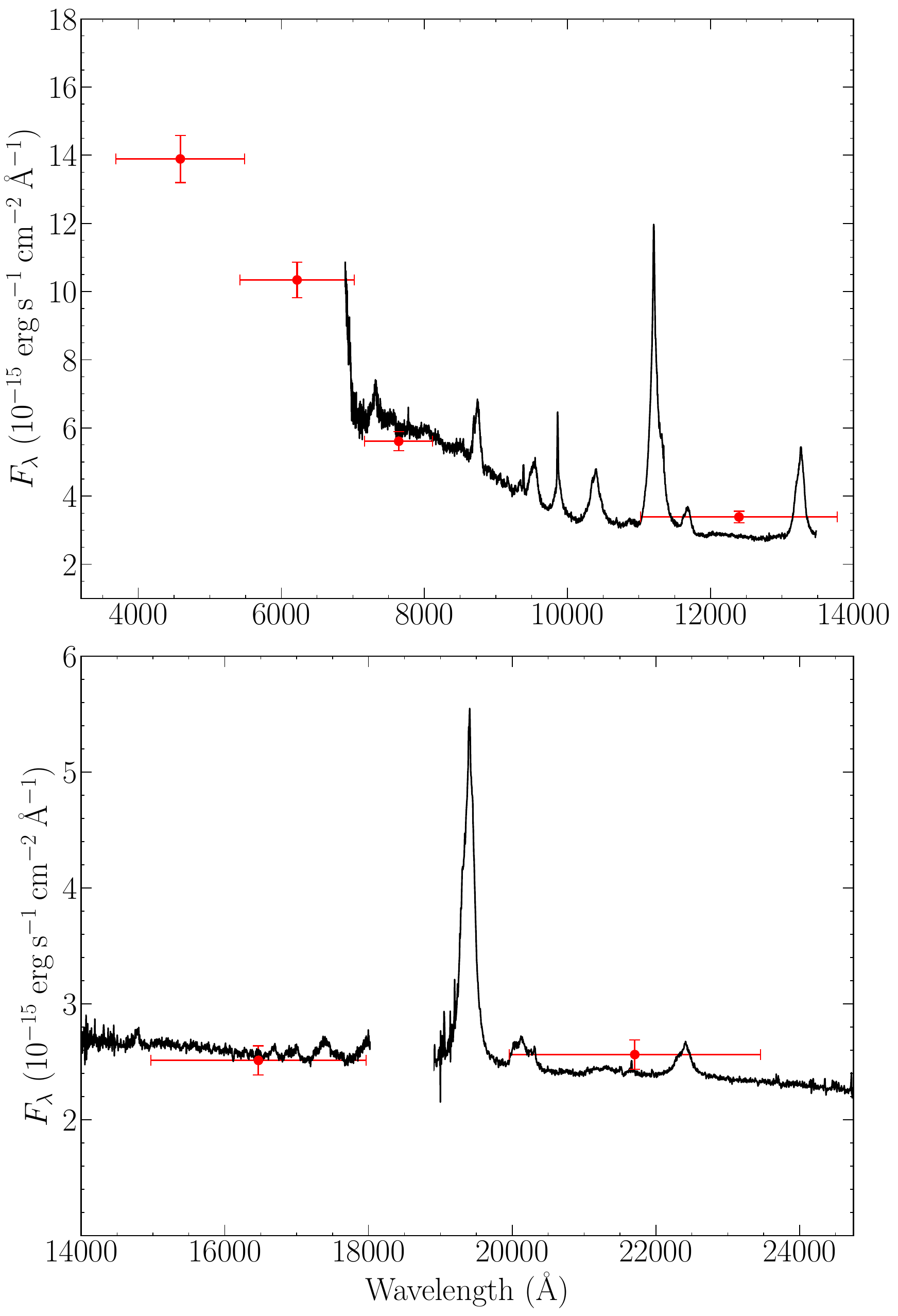}
}
\caption{\label{grondspec} IRTF SpeX near-IR spectrum from 26$\mathrm{^{th}}$ August 2019, highlighted in Figure \ref{z_check} (black), overlaid with the optical and near-IR nuclear fluxes from the GROND photometry obtained on the same day (red filled circles).}
\end{figure}

In Table \ref{grondphot}, we list the optical and near-IR nuclear fluxes. They were obtained with \galfit~\citep{galfit}, a software package that models the object's surface brightness profile with a point spread function (PSF), for the unresolved AGN, and a host galaxy component. We modelled the host galaxy with only a bulge due to the relatively low S/N of the images. In addition, \galfit~simultaneously models the PSF fluxes of the reference stars, which we used to flux-calibrate the optical and near-IR images based on the known stellar fluxes as listed in the SDSS \citep{SloanDR7} and 2MASS \citep{2MASS}. We note that we had to omit the $z$-band images due to lack of SDSS reference stars in this area and that the host galaxy parameters in the $H$ and $K$~bands had to be constrained (based on the other images) rather than left to vary. Figure \ref{grondspec} shows clearly that the GROND photometry matches the spectral shape and flux level of our near-IR spectrum, which provides a useful check of our telluric and correction, and indicates the lack of host galaxy contribution to our near-IR spectra. 

\subsection{Photometric corrections to the spectra} \label{rescaling}

\begin{figure}
\centerline{
\includegraphics[ scale=0.32]{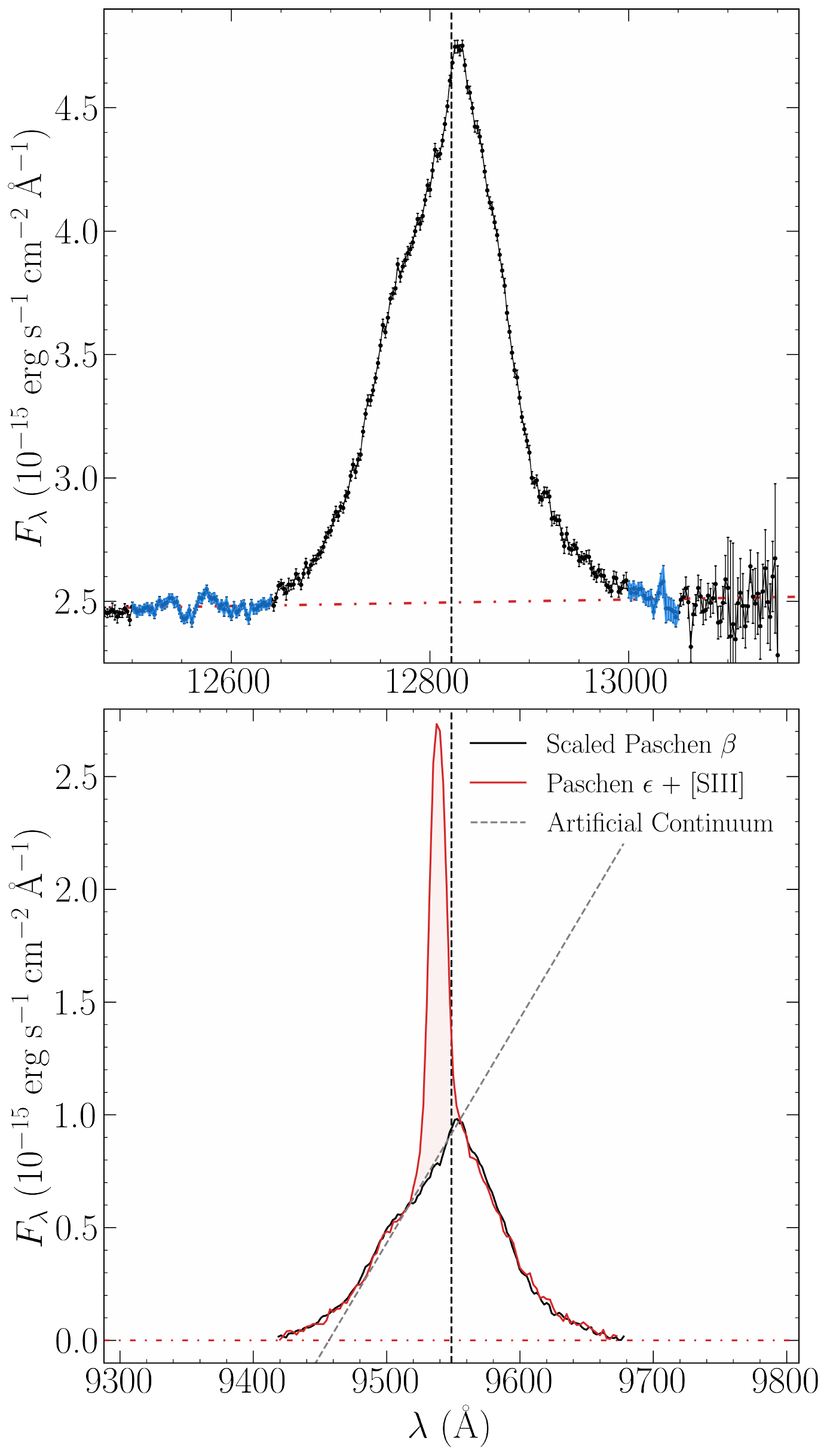}
}
\caption{\label{siii_flux_window}
 \textbf{Top Panel:}  Pa$\beta$ emission line from Mrk 509 spectrum taken on 26 August 2019. Red line represents a linearly fitted continuum, subtracted in order to calculate enclosed flux. The continuum has been fit using a blueward clean continuum window of (12500-12640) \AA{} and the redward clean continuum window of (13000-13050) \AA, these regions are highlighted in blue. The redward window is necessarily short as beyond this there is significant telluric absorption. \textbf{Bottom Panel:}  Mean spectrum of Mrk 509 from all 4 nights with photometric conditions. Pa$\mathrm{\beta}$ profile (black) scaled to match Pa$\mathrm{\epsilon}$ (red), in order to isolate the constant \SIII{} $\mathrm{\lambda 9531}$ component from any variable broad line emission. The slope of the Pa$\mathrm{\beta}$ profile in the \SIII{} window is shown as the artificial continuum in grey. Both panels centre on the rest frame wavelengths of 12821.6 $\mathrm{\Angstrom}$ and 9548.6 $\mathrm{\Angstrom}$ for Pa$\beta$ and Pa$\epsilon$ respectively and have a matched velocity range.
}
\end{figure}

\begin{figure}
\centerline{
\includegraphics[ scale=0.37]{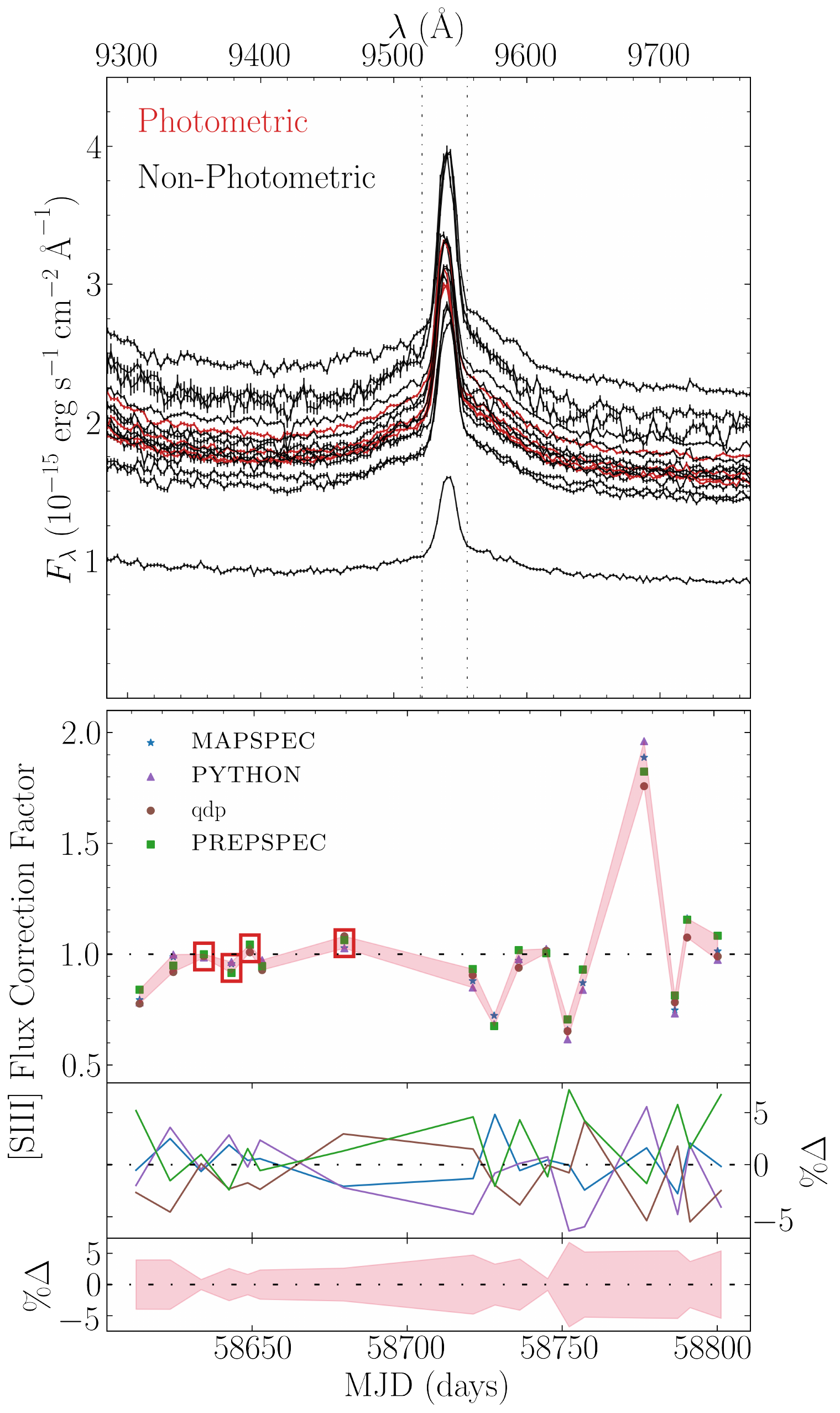}
}
\caption{\label{scale_comp}\textbf{Top Panel:} All 17 Mrk 509 spectra, showing the region around Pa$\mathrm{\epsilon}$ and \SIII. The four photometric spectra are coloured red, the others black.  \textbf{Second Panel:} \SIII{} flux correction factors calculated using four different methods for each night of observation, photometric nights are enclosed by red boxes. The four methods used are as follows; \mapspec{} using spectra with an appended artificial continuum (see Figure \ref{siii_flux_window}), interactive fitting routine qdp with a linear, narrow Gaussian and broad Gaussian component, numerically integrating under the \SIII{} line profile after the removal of the artificial continuum and \prepspec{}. \textbf{Third Panel:} Percentage of the mean correction factor for each value obtained via the four methods described above, coloured similarly to the panel above. \textbf{Bottom Panel:} Percentage range of the four methods around their mean. }
\end{figure}

\begin{figure}
\centerline{
\includegraphics[ scale=0.415]{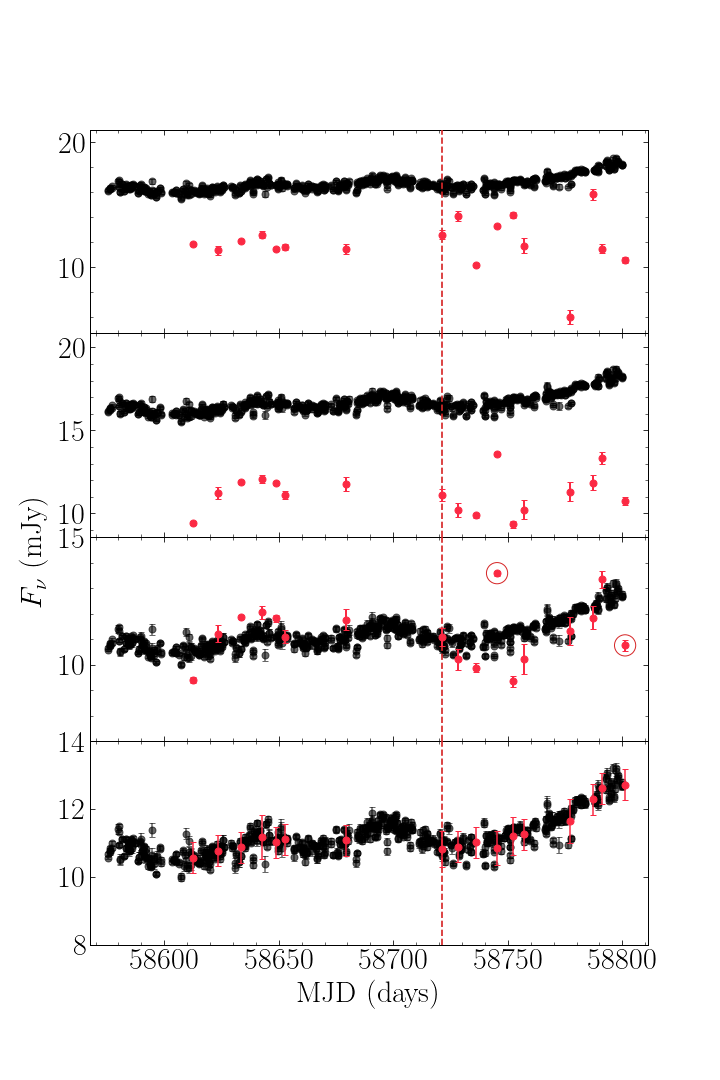}
}
\caption{\label{z_check} \textbf{Top Panel:} LCO $z_s$ band photometry, uncorrected for host galaxy contribution in black, with all 17 spectral LCO $z_s$ band (8180-9220 $\mathrm{\Angstrom}$ observed frame) uncorrected fluxes plotted in red. \textbf{Second Panel:} LCO $z_s$ band photometry, uncorrected for host galaxy contribution in black, with all 17 spectral  LCO $z_s$ band ((8180-9220) $\mathrm{\Angstrom}$ observed frame) fluxes, corrected with \mapspec{} scaling factors are plotted in red.\textbf{Third Panel:} LCO $z_s$ band photometry, corrected for a constant host galaxy contribution in black, with all 17 spectral LCO $z_s$ band (8180-9220 $\mathrm{\Angstrom}$ observed frame) fluxes, corrected with the \mapspec{} scaling factors  plotted in red. Spectral fluxes measured on MJD 58745.30 and 58801.20 are encircled in red as they do not match well with the   LCO $z_s$ band photometry. The night of 26$\mathrm{^{th}}$ August 2019 is highlighted with a dashed red line. \textbf{Bottom Panel:} The \mapspec{} corrected lightcurve has been further corrected to the LCO $z_s$ band photometry by scaling to the weighted mean of all photometric data points within three days of each spectral data point. The errors displayed are defined by the spread of the data to which the spectral points have been corrected, or the mean spread of the photometry within any given 3 day period of the spread if the correction was less than this value. These uncertainties have been propagated to our corrected lightcurves along with the \mapspec{} uncertainties displayed in the panel above.  }
\end{figure}

For the purposes of this work it is essential to distinguish between intrinsic AGN variability and extrinsic sources of variability such as changing night to night observing conditions. Therefore to obtain meaningful lightcurves from the spectra it was necessary to perform a photometric correction to the spectra. Failing to do this could result in a contamination of the final dust response lightcurves \citet{Bentz16}. 

Several software packages employ different approaches to perform the photometric re-scaling of spectra, calibrating out time-dependent instrumental effects, e.g \prepspec{} (written by Keith Horne and discussed in \citet{Shen16}) and \GW{}  \citep{GW92}. This work uses the open source \python{} package \mapspec{} \citep{Faus17} which is an updated and improved version of the \GW{} package. \mapspec{} , like \GW{}, aligns an assumed non variable line profile of a given observed spectrum to a reference spectrum that has been constructed, by applying a smoothing kernel, a flux re-scaling factor, and a wavelength shift. 

In order to correct for time-dependent instrumental effects, a spectral feature which is intrinsically non variable over the baseline of observations is required. Traditionally, for optical campaigns, this is the \OIII{} $\lambda 5007$ narrow emission line. Confirmed via numerous IFU observations \OIII{} $\lambda 5007$ is separated from the ionising source by a pc-to-kpc scale region and therefore constant on timescales much larger than the baseline of observations typically used in reverberation studies such as this \citep{Faus17}. However, this emission line is not visible in the near-IR, therefore we have selected the \SIII{} $\lambda 9531$ narrow emission line, which is analogous to the \OIII{} $\lambda 5007$ region due to its extent.

A complication in the use of the \SIII{} $\lambda 9531$ narrow emission line is that it is blended with the variable broad Pa$\mathrm{\epsilon}$. It was therefore important to separate these two profiles to allow for accurate flux scaling. In order to do this a mean spectrum of our 4 photometric nights was constructed. Using this mean spectrum we isolated and fit the broad Pa$\mathrm{\beta}$  to the broad Pa$\mathrm{\epsilon}$ profile in velocity space and then converted back to wavelength space. This fit is displayed in Figure~\ref{siii_flux_window}.

A linear fit to the Pa$\mathrm{\beta}$ profile within the \SIII{} window of (9520 - 9560) \AA{} was performed to calculate the slope of an artificial continuum which could be used to isolate the intrinsically non-variable \SIII{} narrow line from the variable broad Pa$\mathrm{\epsilon}$ upon which it sits. Figure \ref{scale_comp} top panel shows the Pa$\mathrm{\epsilon}$ and \SIII{} emission region from all 17 Mrk 509 spectra with the real continuum and the broad Pa$\mathrm{\epsilon}$ profile, with the bottom panel displaying the \SIII{} scale factors calculated in four different ways. 

Using \mapspec{}, a reference spectrum was constructed from 4 photometric spectra using the artificial continua as displayed in Figure \ref{siii_flux_window} as the continuum level, thereby isolating the \SIII{} profile. This reference spectrum was used alongside all 17 spectra with artificial continua in order to calculate re-scaling factors.  \mapspec{} allows for three different smoothing methods, a Delta function, a Gaussian and  Gauss-Hermite smoothing kernel. We have taken the final scaling values as the Gauss-Hermite smoothing value as this is the most stable of the three methods \citep{Faus17}. We performed a test using two different continua, the broad line profile and the artificial continua shown in Figure \ref{siii_flux_window}, resulting in an average difference in scale factor of 3.5\% between the two continuum options for the Gauss-Hermite method compared to an average difference in scale factor of 10.4\% for the delta function, therefore confirming this as the most robust smoothing method. 

To quantify the uncertainties from the scaling procedure we have calculated scaling factors using three alternative methods. Firstly by calculating the integrated flux in the \SIII{} line using the artificial continua and scaling these areas to the mean area of all four photometric nights. Secondly, using the interactive fitting package qdp, we modelled the \SIII{} region with a linear continuum plus broad and narrow Gaussian components. The linear continuum and broad line components were subtracted and the flux enclosed by the narrow line was calculated and scaled to the mean of the photometric nights. Finally we used \prepspec{} to calculate a set of flux correction factors. In Figure \ref{scale_comp}, data points representing the photometric nights (enclosed in red) show very little spread in correction factor and all lie within a few percent of unity, as expected. Across all observations, the corrections factors for each method of calculation described above agree within $\lesssim5\%$. Figure \ref{scale_comp} shows that on average \mapspec{} is closest to the mean correction factor for each observation, but the distribution around the mean for the other three correction factors appears random, indicating that one method alone does not significantly increase the spread displayed in the bottom panel.

The two central panels of Figure~\ref{z_check} show the \mapspec{} corrected LCO $z_s$-band spectral fluxes compared with LCO $z_s$-band photometry. Whilst an exact agreement with the photometry would not be necessary to confirm the validity of the correction factors, it is clear that there are two measurements which are very discrepant with respect to the contemporaneous photometric measurements. These large discrepancies reduce the confidence in the scaling procedure and the assumptions upon which it is based, perhaps indicating that there is some extension in the \SIII{} region causing variability in this line between observations not due to atmospheric effects. We therefore scale to the LCO $z_s$-band photometry, as shown in the bottom panel of Figure \ref{z_check}. To do this, we bin the photometric data into 3-day windows around each spectral data point and scale to the weighted mean of this value, we then propagate the weighted standard error of this bin forward as the scaling uncertainty, and in any bin where this is less than the average spread of the data in any 3-day period we adopt this value as the uncertainty on the correction. 

To test for potential spatial extension of the \SIII{} emitting region, we extracted spectra from the Integral Field Unit (IFU) image detailed in \citet{Fischer15} (private communication from R. Wilman and T. Fischer), for each position angles (PAs) of the 17 spectra. By running \mapspec{} over this dataset we found that the distribution of the extended emission is such that the addition of this relatively small contribution in each of the PAs we used does not result in variations above a few percent.

\subsection{Variability in the broad Paschen emission lines} \label{pa_section}

\begin{figure}
\centerline{
\includegraphics[scale=0.4, clip=true]{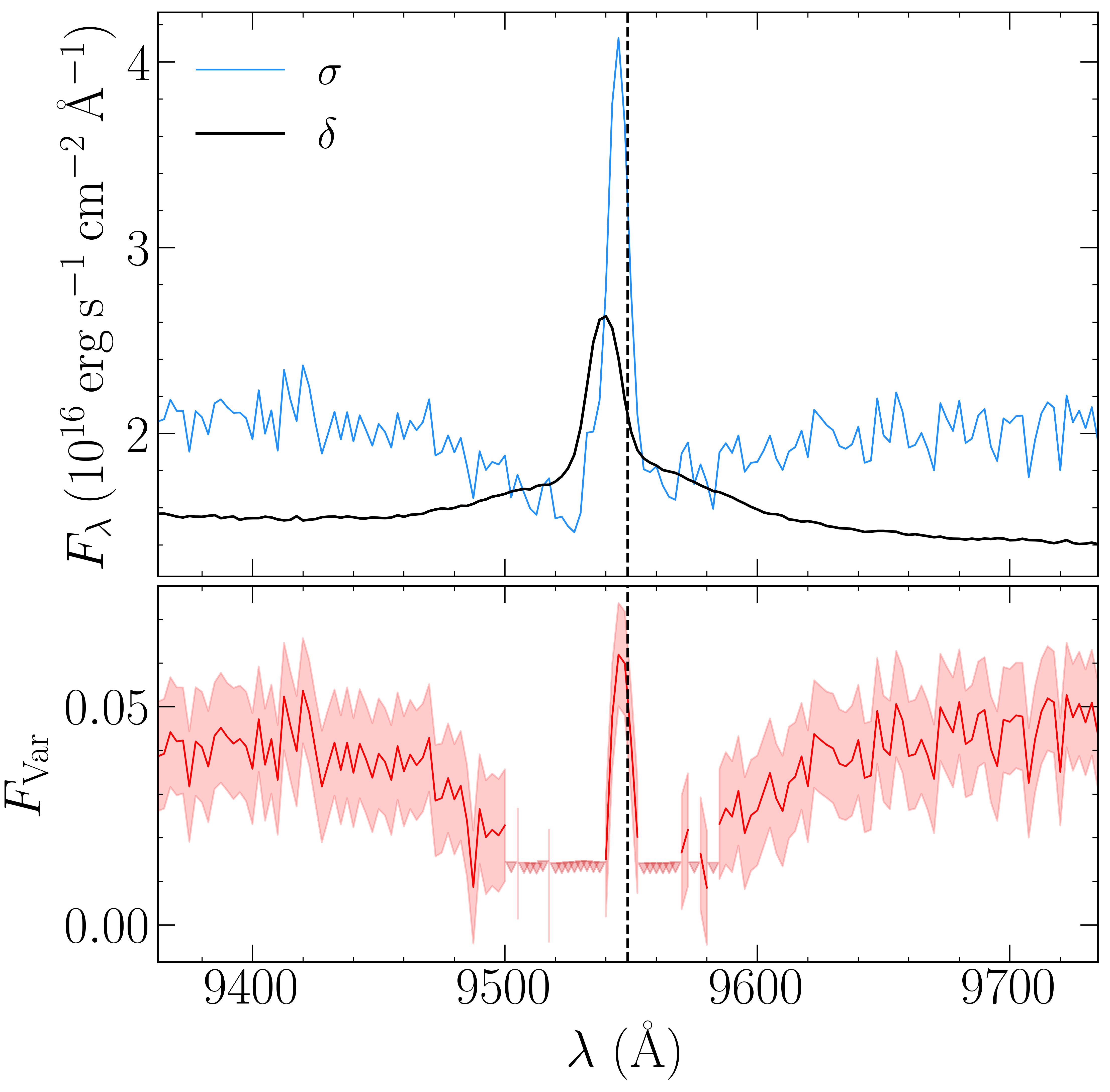}
}
\caption{\label{paschen_fvar_plot} $\mathrm{\delta}$, $\mathrm{\sigma}$ (top panels) and $F_{\mathrm{var}}$ spectrum (bottom panels) for Pa$\mathrm{\epsilon}$ (top plot) and Pa$\mathrm{\beta}$ (bottom plot)  calculated through the process described in Section \ref{pa_section}. $F_{\mathrm{var}}$ values are shown in red with their 1$\sigma$ uncertainties calculated via Monte Carlo methods, represented by the red shaded region, upper limits on $F_{\mathrm{var}}$ are shown as red triangles. $\sigma$ values are displayed as the blue lines along with their 1$\sigma$ uncertainties calculated via Monte Carlo methods. $\delta$ values shown as the black solid line.  }
\end{figure}

As Mrk~509 is a Type~1 Seyfert it exhibits both broad and narrow emission lines in its optical and near-IR spectra. The Paschen series are a group of hydrogen recombination transitions that fall within the spectral range of the IRTF Spex instrument, and therefore are detected in our spectra. Figure \ref{irtfspec} shows these broad emission lines, the most prominent of which are Pa$\mathrm{\epsilon}$, Pa$\mathrm{\beta}$ and Pa$\mathrm{\alpha}$. It is clear from their profile shapes that these emission lines are emitted from the BLR, which is observed to vary due to its illumination by the accretion disc \citep{Pet04}. Therefore it should be possible to measure variability in the Paschen lines from our spectra, and to search for a reverberation signal from our `driving' optical LCO lightcurve and thereby place constraints on the location of the Paschen BLR. 

To do this we have selected the Pa$\mathrm{\epsilon}$ and Pa$\mathrm{\beta}$ lines, opting to reject Pa$\mathrm{\alpha}$ due to the effect of telluric absorption on its blueward wing, which can be seen in Figure \ref{irtfspec}, but is far more prominent on nights of poorer quality. Pa$\mathrm{\epsilon}$ is surrounded by a relatively clean continuum, however there is a slight complication as on top of its broad profile sits \SIII{}, procedures for the removal of Pa$\mathrm{\epsilon}$ from the \SIII{} are discussed in Section \ref{rescaling}. This technique can be used to isolate the Pa$\mathrm{\epsilon}$ flux, and is discussed in more detail in Section \ref{linelcurvesec}.

The root mean square (rms) fractional variation ($F_{\mathrm{var}}$) is a measure of fractional "excess variance", and therefore is a measure of intrinsic variability above that due to flux measurement errors. $F_{\mathrm{var}}$ is calculated as,

\begin{equation}
    F_{\mathrm{var}} = \frac{ ( \sigma^{2} - \delta^2 )^{1/2} }{\langle F \rangle},
    \label{fvar_eq}
\end{equation}

\noindent following \citet{rod-pa97}, where $\langle F \rangle$ is the sample mean of the $N=17$ flux data $F_{i}$;

\begin{equation}
    \langle F \rangle = \frac{1}{N} \sum_{i=1}^{N}F_{i} ,
\end{equation}

\noindent and $\sigma$ is the square root of the sample variance, calculated as,

\begin{equation}
    \sigma^{2} = \frac{1}{N-1} \sum^{N}_{i=1}(F_{i} - \langle F \rangle)^{2} . \label{sigma_eq}
\end{equation}

\noindent Finally, $\delta^2$ is the mean square uncertainty of the fluxes, calculated as,

\begin{equation}
    \delta^2 = \frac{1}{N} \sum_{i=1}^{N} \delta_{i}^{2} , \label{delta_eq}
\end{equation}

\noindent where $\delta_{i}$ is the uncertainty on each flux measurement $F_{i}$.

Figure \ref{paschen_fvar_plot} shows the $F_{\mathrm{var}}$ spectra (calculated as detailed in Equation \ref{fvar_eq}) for the Pa$\mathrm{\epsilon}$ region. Firstly we plot the $\delta$ and $\sigma$ values, defined in Equations \ref{delta_eq} and \ref{sigma_eq} respectively, as a function of wavelength. Secondly the $F_\mathrm{var}$ values are plotted along with their $1\sigma$ uncertainties. We compute and plot an upper limit for any spectral element for which the $\delta$ value exceeds the $\sigma$ value.

We note that applying a percentage error to each flux element of the spectra leads to an apparent dip in the broad line region of the $F_\mathrm{var}$ spectrum. This artefact is resultant from the increase in the absolute error value for spectral elements of higher flux values. This produces a bias in the weighting step as the $\mathrm{\delta}$ values increase (black line Figure~\ref{paschen_fvar_plot}) whereas the $\mathrm{\sigma}$ values do not (blue line Figure~\ref{paschen_fvar_plot}). Therefore spectral elements with higher flux will be weighted more heavily when calculating $F_{\mathrm{var}}$ thus resulting in a dip in the broad line profile $F_{\mathrm{var}}$ values. 

However, in the case of Pa$\mathrm{\epsilon}$ the \SIII{} narrow line profile shows an $F_{\mathrm{var}}$ amplitude of $\sim6\%$. This indicates that when scaled to the photometry, the \SIII{} profile is variable. As the $F_\mathrm{var}$ has been calculated for each spectral element, the \SIII{} profile calculation does include the continuum variability which would increase $F_\mathrm{var}$, but also includes the systematically increased errors which would act to reduce $F_\mathrm{var}$. Despite these competing effects, we still detect variability in the \SIII{} line profile above the level we detect for the continuum.

Whilst the narrow-line region has been found to vary over large timescales, it is unlikely that we are sensitive to this over the 183 days of our observations. The narrow-line region has been found to be extended over large physical size scales, meaning that we could see variation in the \SIII{} region due to our chosen slit width and observational effects such as seeing \citep{pet13}. For reasons discussed in Section~\ref{rescaling} we believe scaling to the photometry to be more reliable than assuming a non-variable \SIII{} component.

\subsection{The spectral continuum components} \label{components}

\begin{table*}
\caption{\label{lumradiustab} 
Physical parameters for the calculation of luminosity-weighted dust radii}
\begin{tabular}{lclcclcclcc}
\hline
Observation & accretion & \multicolumn{3}{c}{blackbody} & \multicolumn{3}{c}{silicate dust} & \multicolumn{3}{c}{carbon dust} \\
Date & disc & \multicolumn{3}{c}{($\beta=0$)} & \multicolumn{3}{c}{($\beta=-1$)} & \multicolumn{3}{c}{($\beta=-2$)} \\
& log~$L_{\rm uv}$ & $T_{\rm d}$ & log~$L_{\rm d}$ & $R_{\rm d,lum}$ & $T_{\rm d}$ & log~$L_{\rm d}$ & $R_{\rm d,lum}$ & $T_{\rm d}$ & log~$L_{\rm d}$ & $R_{\rm d,lum}$ \\
& (erg/s) & (K) & (erg/s) & (lt-days) & (K) & (erg/s) & (lt-days) & (K) & (erg/s) & (lt-days) \\
(1) & (2) & (3) & (4) & (5) & (6) & (7) & (8) & (9) & (10) & (11) \\
\hline
2019 May 09 & 45.28 & 1369$\pm$11 & 44.66 & 168 & 1170$\pm$8  & 43.78 & 1591 & 1022$\pm$6 & 44.52 & 1022 \\
2019 May 20 & 45.33 & 1386$\pm$14 & 44.64 & 174 & 1182$\pm$10 & 43.72 & 1652 & 1031$\pm$8 & 43.94 & 1063 \\
2019 May 30 & 45.36 & 1362$\pm$12 & 44.65 & 187 & 1164$\pm$9  & 43.68 & 1763 & 1018$\pm$6 & 44.05 & 1129 \\
2019 Jun 08 & 45.35 & 1362$\pm$8  & 44.67 & 184 & 1167$\pm$6  & 43.69 & 1734 & 1022$\pm$5 & 44.22 & 1107 \\
2019 Jun 14 & 45.36 & 1352$\pm$10 & 44.65 & 189 & 1158$\pm$8  & 43.68 & 1781 & 1013$\pm$6 & 44.44 & 1140 \\
2019 Jun 18 & 45.33 & 1363$\pm$12 & 44.66 & 180 & 1164$\pm$9  & 43.68 & 1703 & 1017$\pm$6 & 43.97 & 1093 \\
2019 Jul 15 & 45.36 & 1357$\pm$13 & 44.66 & 188 & 1160$\pm$9  & 43.70 & 1775 & 1014$\pm$7 & 43.87 & 1138 \\
2019 Aug 26 & 45.35 & 1353$\pm$14 & 44.69 & 187 & 1157$\pm$10 & 43.68 & 1764 & 1011$\pm$8 & 44.04 & 1132 \\
2019 Sep 02 & 45.25 & 1381$\pm$15 & 44.61 & 160 & 1180$\pm$10 & 43.70 & 1511 & 1030$\pm$8 & 44.91 &  971 \\
2019 Sep 10 & 45.32 & 1367$\pm$15 & 44.70 & 177 & 1167$\pm$10 & 43.72 & 1675 & 1018$\pm$9 & 44.04 & 1078 \\
2019 Sep 19 & 45.41 & 1376$\pm$12 & 44.72 & 194 & 1173$\pm$8  & 43.73 & 1839 & 1022$\pm$7 & 43.98 & 1187 \\
2019 Sep 26 & 45.33 & 1396$\pm$16 & 44.69 & 172 & 1190$\pm$11 & 43.75 & 1629 & 1038$\pm$8 & 44.07 & 1049 \\
2019 Oct 01 & 45.35 & 1365$\pm$14 & 44.69 & 184 & 1166$\pm$10 & 43.70 & 1737 & 1018$\pm$7 & 44.04 & 1116 \\
2019 Oct 21 & 45.38 & 1374$\pm$14 & 44.68 & 188 & 1171$\pm$10 & 43.76 & 1782 & 1021$\pm$8 & 44.33 & 1149 \\
2019 Oct 31 & 45.47 & 1354$\pm$11 & 44.69 & 214 & 1156$\pm$9  & 43.70 & 2029 & 1010$\pm$6 & 43.92 & 1302 \\
2019 Nov 04 & 45.47 & 1316$\pm$14 & 44.72 & 227 & 1127$\pm$10 & 43.70 & 2134 & 986$\pm$8  & 44.02 & 1366 \\
2019 Nov 14 & 45.41 & 1364$\pm$14 & 44.63 & 197 & 1165$\pm$11 & 43.69 & 1864 & 1018$\pm$7 & 44.49 & 1196 \\

\hline
\end{tabular} 

\parbox[]{18cm}{The columns are: (1) Universal Time (UT) date of observation; (2) total accretion disc luminosity; for a blackbody emissivity (3) dust temperature; (4) total dust luminosity and (5) dust radius; for an emissivity law appropriate for silicate dust with small grain sizes of $a \la 0.1~\mu$m (6) dust temperature; (7) total dust luminosity and (8) dust radius; for an emissivity law appropriate for carbon dust with small grain sizes of $a \la 0.1~\mu$m (9) dust temperature; (10) total dust luminosity and (11) dust radius.}

\end{table*}

We aim to compare response- and luminosity-weighted dust radii. For the dust time delay we assembled the lightcurves of both the irradiating flux and hot dust. For the calculation of the luminosity-weighted dust radius we measured the dust temperature and estimated the total accretion disc luminosity that heats the dust. The large wavelength coverage of our cross-dispersed near-IR spectra gives in Mrk~509 roughly half of the hot dust SED, and the improved wavelength coverage of SpeX in the blue samples a considerable part of the accretion disc spectrum, which is expected to dominate the total continuum flux up to a rest-frame wavelength of $\sim 1~\mu$m \citep{L11b, L11a}. We note that, since we used a relatively small spectral aperture of 0.3" which corresponds to a physical size of $\sim$205 pc at the source, the contribution from the host galaxy to the total observed continuum flux is expected to be small in this luminous AGN.

Following \citet{L19}, we decomposed the spectral continuum into two components. We first approximated the rest-frame wavelength range of $\la 1~\mu$m with an accretion disc spectrum, which we subsequently subtracted from the total spectrum. For its calculation we adopted a black hole mass of $M_{\rm BH}=1.09 \times 10^8$~solar masses, which is the value derived by optical reverberation campaigns \citep{Pet04, Bentz15} using a geometrical scaling factor of $f = 4.3$ to convert the measured virial product \citep{Grier13b}. Furthermore, we assumed that the disc extends to $r_{\rm out} = 10^4 r_{\rm g}$, where $r_{\rm g} = G M_{\rm BH}/c^2$ is the gravitational radius, with $G$ the gravitational constant and $c$ the speed of light, the value of $\dot{m}$ was resultant from each fit and an efficiency of $\eta = 0.057$ was assumed. We then fitted the resultant hot dust spectrum at wavelengths $>1~\mu$m with a blackbody, representing emission by large dust grains, and with two blackbodies modified by a power-law of the form $Q_{\lambda} (a) \propto \lambda^{\beta}$, approximating with $\beta=-1$ and $\beta=-2$ the emissivity of sub-micron silicate and carbon dust grains, respectively \citep[][see their Figure 8]{L19}.  Despite our fitting of only two components, there is evidence to suggest that at 1$\mathrm{\mu m}$ diffuse BLR continuum emission could account for 25$\%$ of the total continuum emission with this increasing to 35$\%$ at 8000$\Angstrom$ \citep{Korista19}. This could have an effect on our spectral decomposition, and on our spectral lightcurves, particularly those extracted at wavelengths <1$\mathrm{\mu m}$, see Section \ref{lcurvesec}. Table \ref{lumradiustab} lists the relevant physical parameters extracted from the spectral decomposition. We obtain average temperatures of \mbox{$\langle T \rangle = 1365\pm4$~K}, \mbox{1166$\pm$3~K} and \mbox{1018$\pm$3~K} for emissivity laws with $\beta=0$, $-1$ and $-2$, respectively. 

As in \citet{L19}, we calculated luminosity-weighted dust radii, $R_{\rm d,lum}$, from the best-fit dust temperatures assuming radiative equilibrium between the luminosity of the irradiating source and the dust:

\begin{equation}
\label{Stefan-Boltz}
\frac{L_{\rm uv}}{4 \pi R_{\rm d,lum}^2} = 4 \sigma T^4 \langle Q^{\rm em} \rangle,
\end{equation}

\noindent

where $\sigma$ is the Stefan-Boltzmann constant and $\langle Q^{\rm em} \rangle$ is the Planck-averaged value of $Q_{\lambda} (a)$. We approximated $L_{\rm uv}$ with the accretion disc luminosity and have used for the Planck-averaged emission efficiencies in the case of \mbox{$\beta=-1$} a value of \mbox{$\langle Q^{\rm em} \rangle=0.0210$} appropriate for silicates of \mbox{$T=1259$~K} and \mbox{$a=0.1~\mu$m} \citep{Laor93} and in the case of \mbox{$\beta=-2$} a value of \mbox{$\langle Q^{\rm em} \rangle=0.0875$} appropriate for graphite of \mbox{$T=1000$~K} and \mbox{$a=0.1~\mu$m} \citep{Draine16}. The average luminosity-weighted dust radii are \mbox{$\langle R_{\rm d,lum} \rangle = 186\pm4$~light-days}, \mbox{1763$\pm$36~light-days} and \mbox{1132$\pm$23~light-days} in the case of a blackbody, and small-grain silicate and carbon dust, respectively.

\section{Spectral and emission line lightcurves} \label{reverberation}

\subsection{The observed spectral continuum light-curves} \label{lcurvesec}

\begin{figure}
\centerline{
\includegraphics[ scale=0.39]{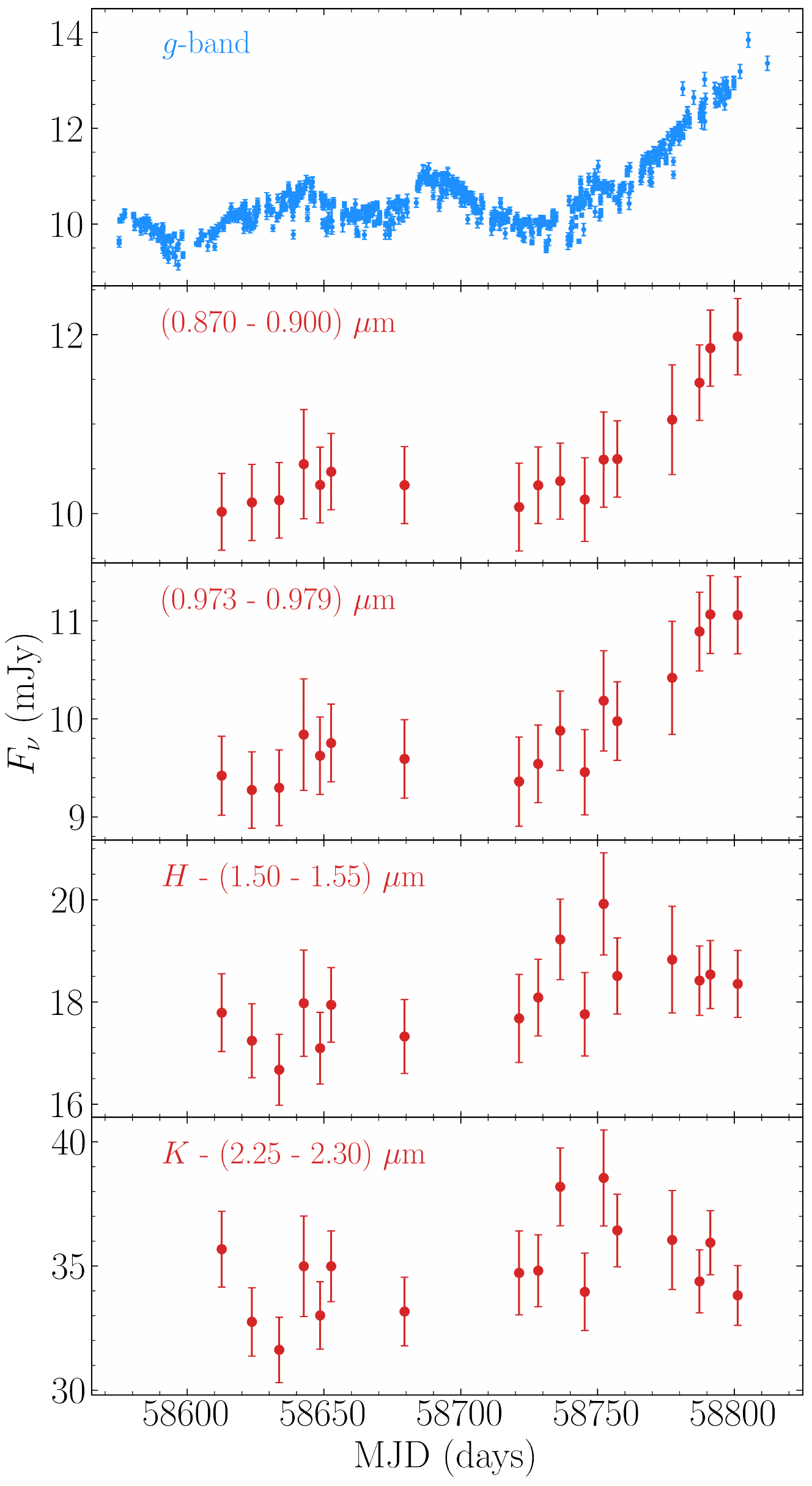}
}
\caption{\label{spec_lc_corr} Top-bottom, LCO $g$-band, (0.870-0.900) $\mathrm{\mu m}$, (0.973-0.979) $\mathrm{\mu m}$, \textit{H}-band (1.50-1.55) $\mathrm{\mu m}$ and \textit{K}-band (2.25-2.30) $\mathrm{\mu m}$ spectral lightcurves. Fluxes have been calibrated onto an absolute scale using \mapspec{} scale factors and then a secondary photometric correction, therefore any variability is considered to be intrinsic.  }
\end{figure}

Lightcurves for four spectral continuum windows are constructed from fluxes extracted from near-IR spectra and displayed in Figure~\ref{spec_lc_corr}. In addition, we have the $g$-band photometric light-curve for the accretion disc, shown in Figure~\ref{ztf_lco}. Ideally, we would like to measure the accretion disc flux at the shortest wavelengths, where it is least contaminated by the hot dust, however, the S/N~ratio of the near-IR spectra decreases significantly towards short wavelengths. Therefore, we measured the accretion disc flux in the 60~\AA~wide rest-frame wavelength region of $\lambda=9730-9790$~\AA, which lies between the two broad hydrogen emission lines Pa$\epsilon$ and Pa$\delta$ and is known to be line-free, and also in the wider (300~\AA) rest-frame wavelength region of $\lambda=8700-9000$~\AA, which is not contaminated by significant emission-line flux in Mrk~509 (see Figure~\ref{irtfspec}).

We measured the  flux in two line-free, 500~\AA~wide rest-frame wavelength regions, namely, $\lambda=2.25-2.30~\mu$m, which is towards the red end of the $K$~band and thus close to the peak of the blackbody spectrum, and $\lambda=1.50-1.55~\mu$m, which is close to the middle of the $H$~band. These bands were selected due to the absence of any emission line contamination and the lack of any telluric effects even in the worst quality spectra. These two wavebands have a significantly lower contribution from the accretion disc than the two shorter wavelengths selected. In Figure \ref{spec_lc_corr} the 30$\%$ rise in $g$-band flux between MJD $\sim$57830 and $\sim$58800 has no clear counterpart in the  \textit{H} and \textit{K} bands as opposed to the shorter-wavelength bands which possess this distinctive rise also seen in our $g$-band lightcurve, thus demonstrating the reduction in accretion disc contribution well. The reduction in accretion disc contribution at longer wavelengths is well documented in Mrk 509 and similar sources \citep{Hern-Caba15,Hern-Caba17,Kura03}. \citet{Hern-Caba17} show that at \textit{H} the contributions from the accretion disc and hot dust are roughly equal, whereas at \textit{K} the hot dust emission significantly dominates over that of the accretion disc. It is worth noting that the shorter wavelength variability that we sample could contain a significant contribution from diffuse BLR continuum emission, or a wind on the inner edge of the BLR, rather than simply being from the disc \citep{Netzer21,Kara21}.

It is clear from Figure~\ref{spec_lc_corr} that the level of detected variability decreases with increasing wavelength, with the most significant variability being detected in the $g$-band.

From the (126$\mathrm{\pm}$11) lt-day lag measured by \citet{Kosh14} and the interferometrically determined radius of (296$\mathrm{\pm}$31) lt-days \citep{gravity}, along with our luminosity weighted dust radius of (186$\mathrm{\pm}$4) lt-days, it appears that we should expect to measure a lag of ~(100-200) days. A time lag of this size would necessarily require a smoothing timescale of a similar order meaning that the hot dust emission originating from the torus would not display a high level of variability on short timescales. This could explain why, with our limited baseline and number of observations, 17 data points across 183 days, we detect little variability in the hot dust emission. We would only expect the already low level of variability seen in the \textit{H} and \textit{K}-band lightcurves to decrease if the accretion disc component were removed, and therefore we would almost certainly not be sensitive to variability in the hot dust emission alone.

\subsection{Emission line light-curves} \label{linelcurvesec}

\begin{figure}
\centerline{
\includegraphics[scale=0.37, clip=true]{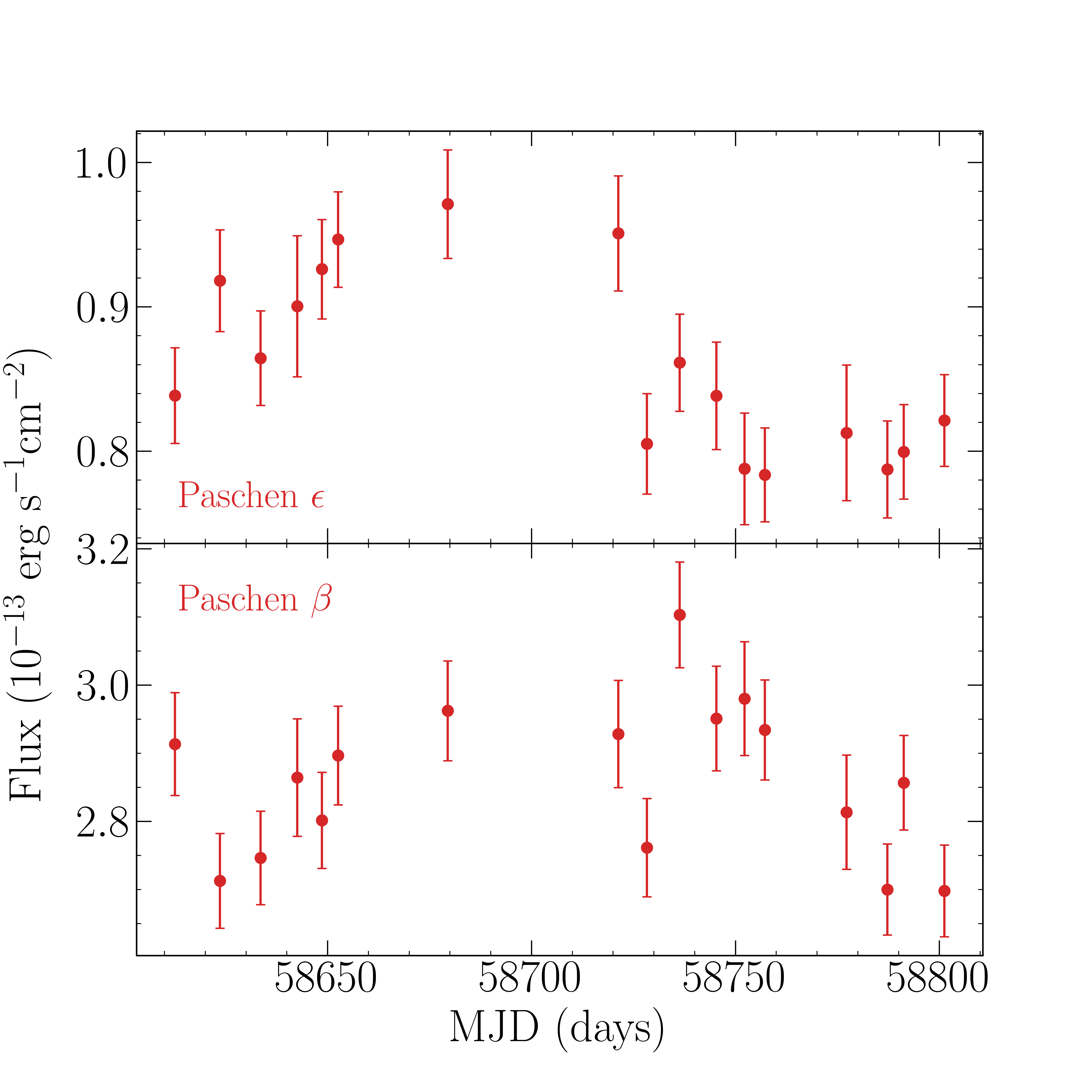}
}
\caption{\label{paschen_lightcurves} Pa$\epsilon$ and  Pa$\beta$ emission line lightcurve, constructed using the procedure described in Section \ref{linelcurvesec}. }
\end{figure}

To construct lightcurves, we extract the Pa$\mathrm{\epsilon}$ and Pa$\mathrm{\beta}$ line fluxes. For Pa$\mathrm{\epsilon}$  we first fit a linear continuum to redward and blueward clean continuum regions, shown as the red dot-dashed line in Figure \ref{siii_flux_window}. We subtract this linear continuum, and then integrate under the data using Simpson's method. We must subtract the \SIII{} component from each observation in order to obtain the flux enclosed by Pa$\mathrm{\epsilon}$. As detailed in Section~\ref{rescaling} we opt to scale to the photometry due to major discrepancies in the scaling procedures relying on the \SIII{} being constant. Therefore the subtracted \SIII{} values vary by $\sim6\%$ in line with our calibration procedure. The flux extraction process was slightly more straightforward for the Pa$\mathrm{\beta}$ profile: we fit a linear continuum to relatively clean blueward and redward continuum regions, again displayed by the dashed line in Figure~\ref{siii_flux_window}. The selected redward continuum window is necessarily short in this case as this region is affected by telluric absorption.  We then subtract this continuum and simply integrate under the line, using Simpson's method, to give the fluxes displayed in Figure~\ref{paschen_lightcurves}. The Paschen line flux uncertainties shown in Figure~\ref{paschen_lightcurves} were calculated using the Monte-Carlo method, with 1000 iterations, selecting the 1$\sigma$ bounds of the resultant Gaussian distribution and then numerically propagating this with a $2\%$ systematic from the continuum selection. 

 Figure \ref{paschen_lightcurves} shows the lightcurves for Pa$\mathrm{\epsilon}$ and Pa$\mathrm{\beta}$, it is clear that there is not a high level of variability. This is confirmed by the calculated $F_{\mathrm{var}}$ values listed in Table \ref{f_var_paschen}.

\begin{table}
\centering
\caption{\label{f_var_paschen} 
The rms fractional variation ($F_{\mathrm{var}}$) in each relevant waveband, unweighted mean flux ($\langle F \rangle$), sample variance ($\sigma$) and the RMS uncertainty ($\delta$) calculated following \citet{rod-pa97}. }
\begin{tabular}{lcccccr}
\hline
Line &    $F_{\mathrm{var}}$ & $\langle F \rangle$  &$\sigma$& $\delta$ \\
 & & (erg $\mathrm{s^{-1} cm^{-2}}$) & (erg $\mathrm{s^{-1} cm^{-2}}$) &  (erg $\mathrm{s^{-1} cm^{-2}}$) \\
\hline
Pa$\mathrm{\epsilon}$  & 0.060$\mathrm{\pm}$0.012 & 0.86 & 0.06 & 0.04\\
Pa$\mathrm{\beta}$ &  0.030$\mathrm{\pm}$0.007 & 2.860 & 0.11 & 0.07 \\

\hline
\end{tabular}

\end{table}

\section{Response-weighted dust and emission line radii} \label{reverberation} \label{revresults}

\subsection{Continuum and hot dust} \label{spec_jav}

\javelin{} is an open source reverberation mapping package \citep{JAV_11}. Through the use of a Bayesian framework alongside an MCMC sampler, \javelin{} creates a statistical model of a `driving' lightcurve and then assumes `response' lightcurves can be modelled as smoothed, scaled and shifted versions of this. The main assumption upon which \javelin{} relies is that AGN variability is stochastic, and therefore can be accurately modelled as a Damped Random Walk (DRW). 

\begin{figure*} 
\centerline{
\includegraphics[scale=0.28, clip=true]{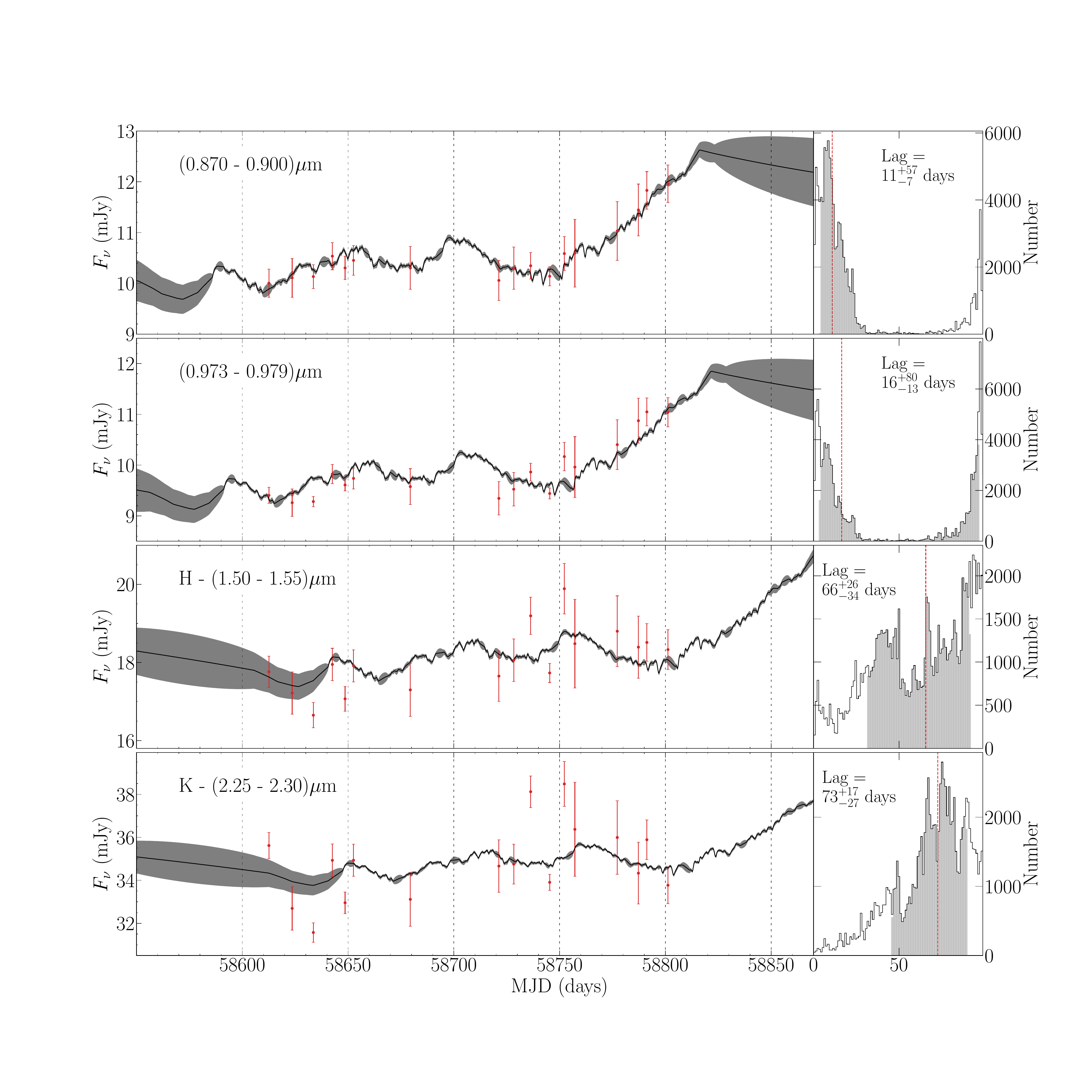}
}
\caption{\label{zzhk_jav} \javelin{} results for the four spectroscopic bands, (0.870-0.900) $\mathrm{\mu m}$, (0.973-0.979) $\mathrm{\mu m}$, (1.50-1.55) $\mathrm{\mu m}$, (2.25-2.30) $\mathrm{\mu m}$, displayed from top to bottom. Each run consists of $10^{5}$ iterations, with the determined lags displayed on the right hand of each panel. The 17 red data points represent spectral data points, and the continuous grey curve is the smoothed, shifted and scaled \javelin{} model of the `driving' $g$-band lightcurve displayed in Figure \ref{ztf_lco}. Lag limits placed at (0-100) days.
}
\end{figure*}

Utilising the LCO and ZTF $g$-band data as a `driving' lightcurve, and the four spectroscopic near-IR lightcurves shown in Figure~\ref{spec_lc_corr} as  `response' lightcurves we run \javelin{} with $10^{5}$ iterations and a (0-100) day limit on the time lag search range in the form of strong preferential priors. This lag range was selected in order to prevent a significant number of spectral data points being shifted into a seasonal gap in the `driving' lightcurve.  The results are displayed in Figure~\ref{zzhk_jav}.

Figure~\ref{zzhk_jav} shows that the (0.870-0.900) $\mathrm{\mu m}$ and (0.973-0.979) $\mathrm{\mu m}$ bands clearly reverberate on timescales shorter than $\sim$25 lt-days, whereas the \textit{H} and \textit{K}-band lightcurves tend towards longer lags. This is to be expected given that the two shorter-wavelength bands are dominated by the accretion disc, which would reverberate on much shorter time scales than any hot dust emission which would dominate to a larger degree at \textit{H} and even more so at  \textit{K}. 

Due to the limited sampling of our data and $\sim$150 day seasonal gaps in the `driving' $g$-band lightcurve, there are strong aliasing problems, and we cannot therefore report any confident lag. However we can state that the \textit{H} and \textit{K}-band lightcurves do not reverberate at <40 days. This is consistent with previous findings. \citet{Kosh14} measure a reverberated dust lag in the \textit{K} band of (126.8$\mathrm{\pm}$11) lt-days, and \citet{grav20} measure the spatially resolved torus at a radius of (296 ± 31) lt-days using near-IR interferometry.

We are not only limited by a lack of significant variability features over the majority of our campaign, but also the data gaps present in both the spectral dust, and photometric continuum lightcurves. The large seasonal data gap shown clearly in Figure~\ref{ztf_lco} between MJD $\sim$58450 and $\sim$58550 limits our ability to measure lags from the hot dust. The $K$-band lag measured by \citet{Kosh14} of (126.8$\mathrm{\pm}$11) lt-days would be extremely difficult to measure with our data-set as a lag of this length would place much of the spectral data into a seasonal data gap, and the length of the lag would smooth out what little variability we do sample in the hot dust response. In fact any lag of more than $\sim$65 days would place a significant proportion of our already limited data sample into this seasonal gap, which greatly limits the ability to measure a credible hot dust response with our current sampling.

In Appendix:A we present a wider range of positive and negative lags to demonstrate the prevalence of alias lags in all of the spectral lightcurves.

\subsection{The broad emission line region}\label{pa_rev_section}

\begin{figure*} 
\centerline{
\includegraphics[scale=0.28, clip=true]{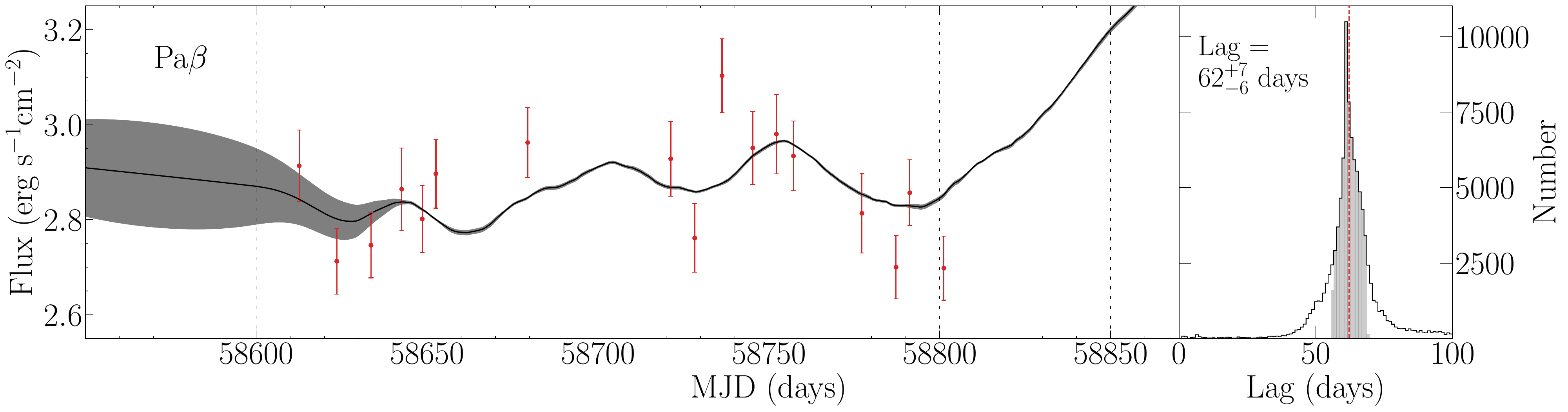}
}
\caption{\label{eps_beta_jav} \javelin{} results for the Pa$\mathrm{\beta}$ broad emission line. Each run consists of $10^{5}$ iterations, with the determined lags displayed on the right hand of each panel. The 17 red data points represent enclosed emission line fluxes, and the continuous grey curve is the smoothed, shifted and scaled \javelin{} model of the `driving' $g$-band lightcurve displayed in Figure \ref{ztf_lco}.  Lag limits placed at (0-100) days.
}
\end{figure*}

\begin{figure*} 
\centerline{
\includegraphics[scale=0.28, clip=true]{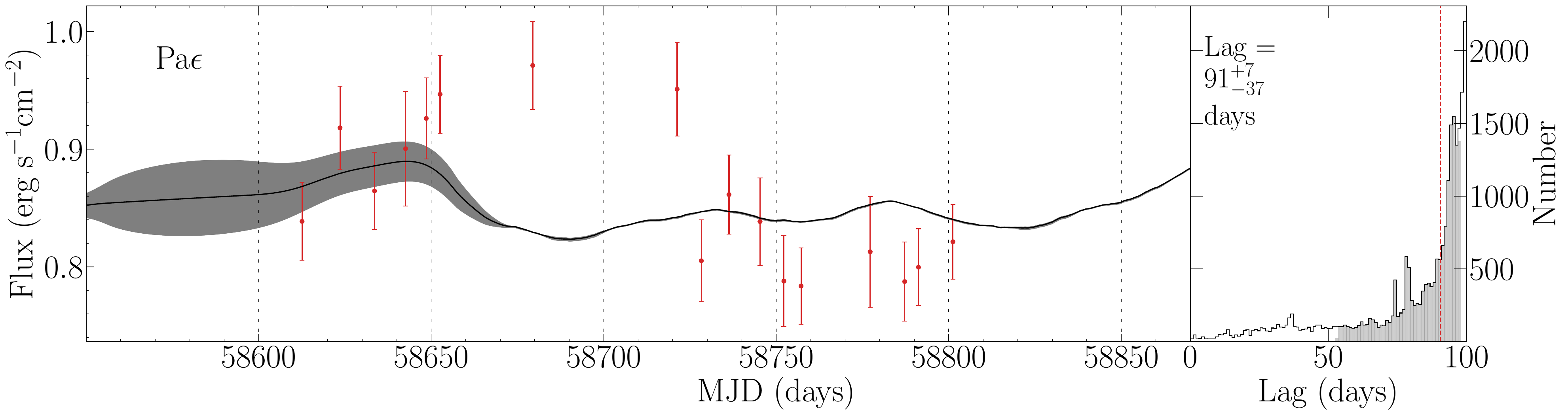}
}
\caption{\label{eps_jav} \javelin{} results for the Pa$\mathrm{\epsilon}$ broad emission line. Each run consists of $10^{5}$ iterations, with the determined lags displayed on the right hand of each panel. The 17 red data points represent enclosed emission line fluxes, and the continuous grey curve is the smoothed, shifted and scaled \javelin{} model of the `driving' $g$-band lightcurve displayed in Figure \ref{ztf_lco}.  Lag limits placed at (0-100) days.
}
\end{figure*}

Literature values for the H$\mathrm{\beta}$ lag in Mrk 509 place the Balmer BLR at $\mathrm{82.3^{+6.3}_{-5.6}}$ days and $\mathrm{79^{+6}_{-7}}$ days for $\tau_{\mathrm{cent}}$ and $\tau_{\mathrm{peak}}$ respectively, using Cross-Correlation Function (CCF) analysis \citep{Pet04}. Unless Mrk 509 has changed state significantly since these measurements were taken (1989-1996), we would expect to see the Paschen line lags to at least be this large. As with measuring reverberation signals in the hot dust of Mrk 509, this presents a problem due to the sparse sampling, short baseline of observations and data gaps present in our data.

As in the case of the hot dust lightcurves, we used \javelin{} to search for Paschen line reverberation signals, the results of which are displayed in Figure \ref{eps_beta_jav}.

\javelin{} reports a lag of $62^{+7}_{-6}$ days in Pa$\mathrm{\beta}$. This result is consistent (within 2$\mathrm{\sigma}$) with previous measurements of the H$\mathrm{\beta}$ lag \citep{Pet04}. Our reverberation result of  $62^{+7}_{-6}$ days, does show some smoothing, with a value of $21^{+29}_{-19}$ days, however this is not as large as we would expect for a lag of this size. In addition, Figure \ref{eps_beta_jav} shows that with a lag of 62 days, 3 out of 17 data points are shifted into a data gap and do not align with any optical data points. Upon visual inspection, we see that the remaining 10 data points do not show a good agreement with the pattern of variability displayed by the optical lightcurve. Therefore we do not report this result with a high level of confidence. However, as with the $H$ and $K$-band spectral lightcurves, we are able to place a lower limit on the Pa$\mathrm{\beta}$ lag of $\sim$40 days as we see no counterpart in the lightcurve of the distinctive rise in flux in the last 80 days of the campaign, marked in red in Figure \ref{ztf_lco}.

We can place this same lower limit on Pa$\mathrm{\epsilon}$ which shows the same distinct absence of the major rise in flux present in our `driving' lightcurve. We measure the Pa$\mathrm{\epsilon}$ at $91^{+7}_{-37}$ days, shown in Figure~\ref{eps_jav}. This lag places 6 of our 17 data points into a data gap and therefore is not reported with a high level of confidence despite being consistent with previous measurements of the H$\mathrm{\beta}$ lag \citep{Pet04}.

\javelin{} results for Pa$\mathrm{\epsilon}$ and Pa$\mathrm{\beta}$ are displayed in separate plots as the analysis was carried out separately. Running \javelin{} separately resulted in a fit with higher level of smoothing in both cases, and therefore in some senses a more physical result.  

In Appendix:A we present a wider range of positive and negative lags to demonstrate the prevalence of alias lags in both Pa$\mathrm{\epsilon}$ and Pa$\mathrm{\beta}$.

\section{Summary and conclusions}

We monitored the AGN Mrk 509 to document its infrared (hot dust) and optical (accretion disc) variations. Our IRTF spectra span 0.7 to 2.4 microns at 17 epochs over 200 days from 2019 May to Nov. The $g$-band monitoring spans 5 seasons with sub-day cadence during the IR campaign.  

Using spectral decomposition we measure luminosity weighted dust radii for different astro-chemical regimes to be \mbox{$\langle R_{\rm d,lum} \rangle = 186\pm4$~light-days}, \mbox{1763$\pm$36~light-days} and \mbox{1132$\pm$23~light-days} in the case of a blackbody, small-grain silicate and carbon dust, respectively. Comparison with the previously measured photometric dust response time of (126$\pm$11) days \citep{Kosh14} and an interferometrically determined radius of ($\mathrm{296\pm31}$) lt-days \citep{grav20} suggests that the dust emits like a black body and therefore is most likely to be composed of large carbonaceous grains. This is consistent with results for the Seyfert NGC 5548 \citep{L19}.

Our $g$-band lightcurve shows a distinctive rise by $\sim$30\% in the last 80 days of the 2019 season, with no clear counterpart in the $H$ and $K$ lightcurves, which have intrinsic variations less than $\sim$4\%. The little variability we do detect in \textit{H} and \textit{K} does not reverberate on timescales less than 40 days and therefore we find a lower limit on the dust response in Mrk 509 which is consistent with previously measured reverberation measurements and also interferometric measurements of the hot dust radius \citep{Kosh14,grav20}. 

We are also limited in our ability to recover a dust lag due to seasonal gaps in both the driving and response lightcurves. This combined with the poor cadence and low levels of variability in the dust response means that application of a large smoothing width would remove any significant variability, which is already low.

We measure lags of $62^{+7}_{-6}$ days and $91^{+7}_{-37}$ days for Pa$\mathrm{\beta}$ and Pa$\mathrm{\epsilon}$ respectively. The Pa$\mathrm{\beta}$ lag is considered a more reliable measurement than that determined for Pa$\mathrm{\epsilon}$ due to a more concentrated peak, fewer data points placed in seasonal gaps of the `driving' lightcurve and a more reasonable smoothing result. The Pa$\mathrm{\epsilon}$ result is consistent with literature values within 1$\sigma$ and the Pa$\mathrm{\beta}$ result consistent within 2$\sigma$ \citep{Pet04}.  However due to the seasonal gaps in our `driving' lightcurve and poor sampling of the emission line lightcurves, we do not report either of these results with a high level of confidence.  As with the $H$ and $K$-bands, we can place a lower limit on the lag of $\sim$40 days as we do not detect any lags below this value and do not recover the distinctive variability feature seen in our `driving' lightcurve during the last $\sim$80 days of our spectral campaign.

A longer baseline of observation, and a higher cadence is needed to recover a reverberated signal from the hot dust and place constraints on the physical size scales of this emission region.

\newpage

\section*{Acknowledgments}
We would like to thank the anonymous referee for their helpful
comments, which improved the manuscript. J.A.J.M acknowledges the support of STFC studentship (ST/S50536/1). MJW acknowledges support from an Leverhulme Emeritus Fellowship, EM-2021-064. K.H. and J.V.H.S. acknowledge support from STFC grant ST/R000824/1. Research by A.J.B. is supported by NSF grant AST-1907290. The authors thank Dr Richard Wilman for the extraction and provision of pseudo-slit spectra from the IFU datacube. This work makes use of observations from the Las Cumbres Observatory global telescope network.  Based on observations obtained with the Samuel Oschin 48-inch Telescope at the Palomar Observatory as part of the Zwicky Transient Facility project. ZTF is supported by the National Science Foundation under Grant No. AST-1440341 and a collaboration including Caltech, IPAC, the Weizmann Institute for Science, the Oskar Klein Center at Stockholm University, the University of Maryland, the University of Washington, Deutsches Elektronen-Synchrotron and Humboldt University, Los Alamos National Laboratories, the TANGO Consortium of Taiwan, the University of Wisconsin at Milwaukee, and Lawrence Berkeley National Laboratories. Operations are conducted by COO, IPAC, and UW. Based on observations obtained with the Samuel Oschin Telescope 48-inch and the 60-inch Telescope at the Palomar
Observatory as part of the Zwicky Transient Facility project. ZTF is supported by the National Science Foundation under Grant
No. AST-2034437 and a collaboration including Caltech, IPAC, the Weizmann Institute for Science, the Oskar Klein Center at
Stockholm University, the University of Maryland, Deutsches Elektronen-Synchrotron and Humboldt University, the TANGO
Consortium of Taiwan, the University of Wisconsin at Milwaukee, Trinity College Dublin, Lawrence Livermore National
Laboratories, and IN2P3, France. Operations are conducted by COO, IPAC, and UW.

\section*{Data Availability}

The data underlying this article will be shared on reasonable request to the corresponding author.

\bibliographystyle{mnras}
\bibliography{references}

\newpage

\appendix \label{app}
\section{\javelin{} - Alias lags}

Here we present \javelin{} results for the spectral and emission line lightcurves shown in Sections~\ref{spec_jav} and \ref{pa_rev_section} respectively. These results are obtained via identical \javelin{} runs apart from the extension of the (0-100) day lag limit to (-800,+600) day lag limits. The results clearly show strong aliasing problems due to seasonal gaps in the optical lightcurve. With the exception of the (0.870-0.900) $\mathrm{\mu m}$ and (0.973-0.979) $\mathrm{\mu m}$ bands which show a very strong peak between 0 and 20 days, the other bands, and emission line lightcurves display an almost symmetric pattern of lag peaks indicating the dominance of alias lags due to seasonal gaps in the optical lightcurve.

\begin{figure*} 
\centerline{
\includegraphics[scale=0.28, clip=true]{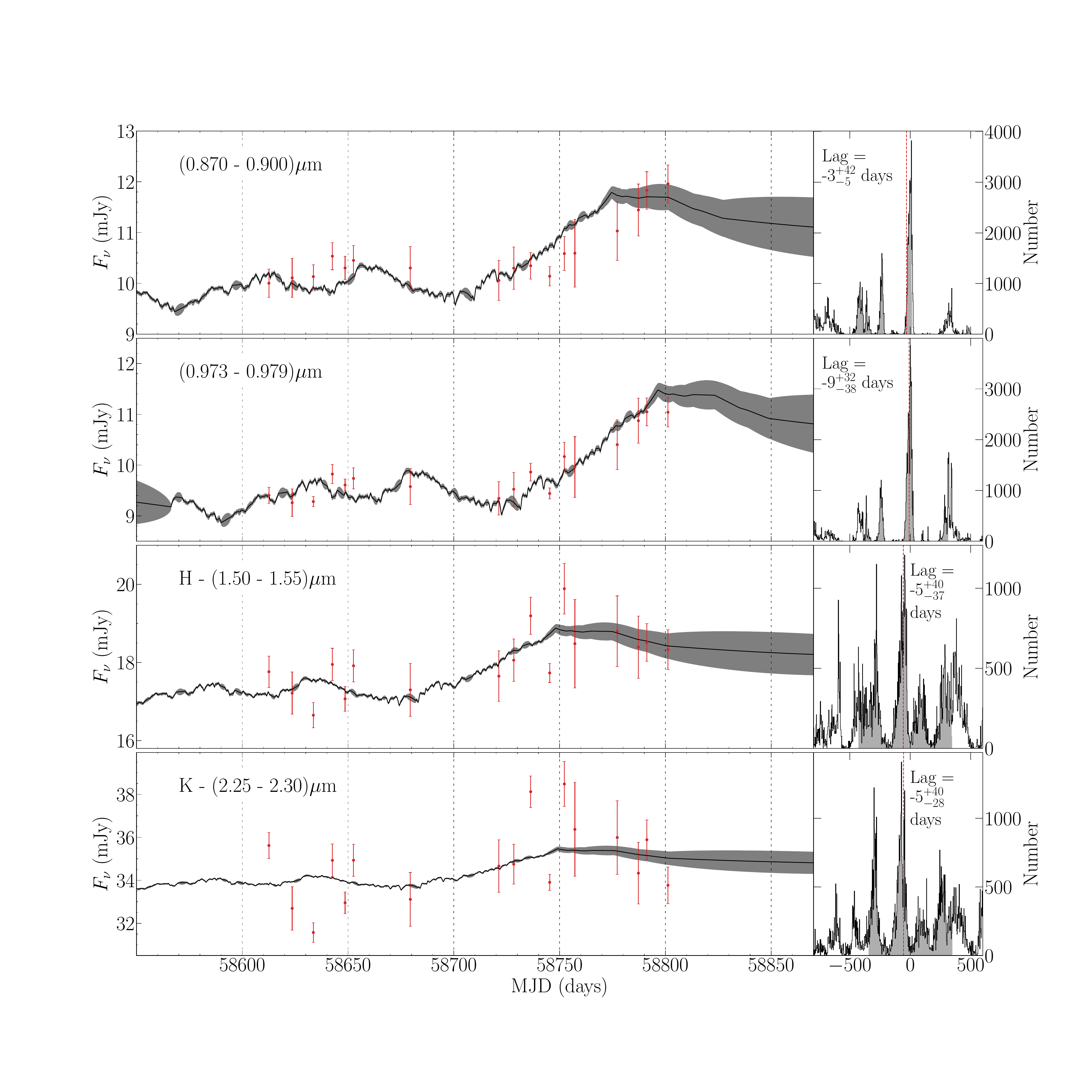}
}
\caption{\label{zzhk_jav_longlim} \javelin{} results for the four spectroscopic bands, (0.870-0.900) $\mathrm{\mu m}$, (0.973-0.979) $\mathrm{\mu m}$, (1.50-1.55) $\mathrm{\mu m}$, (2.25-2.30) $\mathrm{\mu m}$, displayed from top to bottom. Each run consists of $10^{5}$ iterations, with the determined lags displayed on the right hand of each panel. The 17 red data points represent spectral data points, and the continuous grey curve is the smoothed, shifted and scaled \javelin{} model of the `driving' $g$-band lightcurve displayed in Figure \ref{ztf_lco}.  Lag limits placed at (-800 - +600) days.
}
\end{figure*}

\begin{figure*} 
\centerline{
\includegraphics[scale=0.28, clip=true]{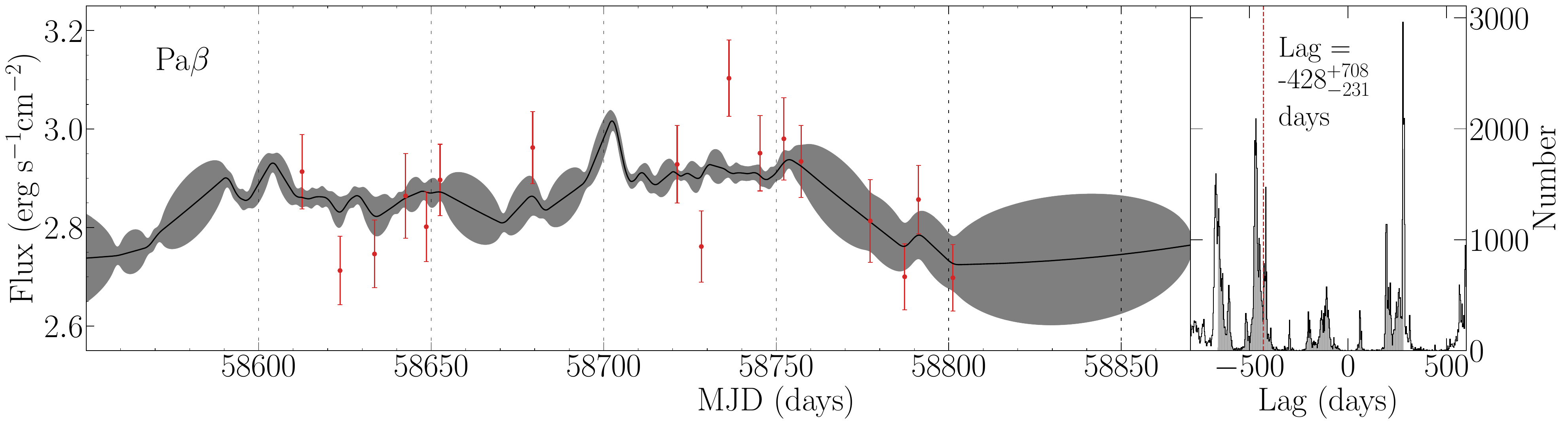}
}
\caption{\label{beta_jav_longlim} \javelin{} results for the Pa$\mathrm{\beta}$ broad emission line. Each run consists of $10^{5}$ iterations, with the determined lags displayed on the right hand of each panel. The 17 red data points represent enclosed emission line fluxes, and the continuous grey curve is the smoothed, shifted and scaled \javelin{} model of the `driving' $g$-band lightcurve displayed in Figure \ref{ztf_lco}. Lag limits placed at (-800 - +600) days.
}
\end{figure*}

\begin{figure*} 
\centerline{
\includegraphics[scale=0.28, clip=true]{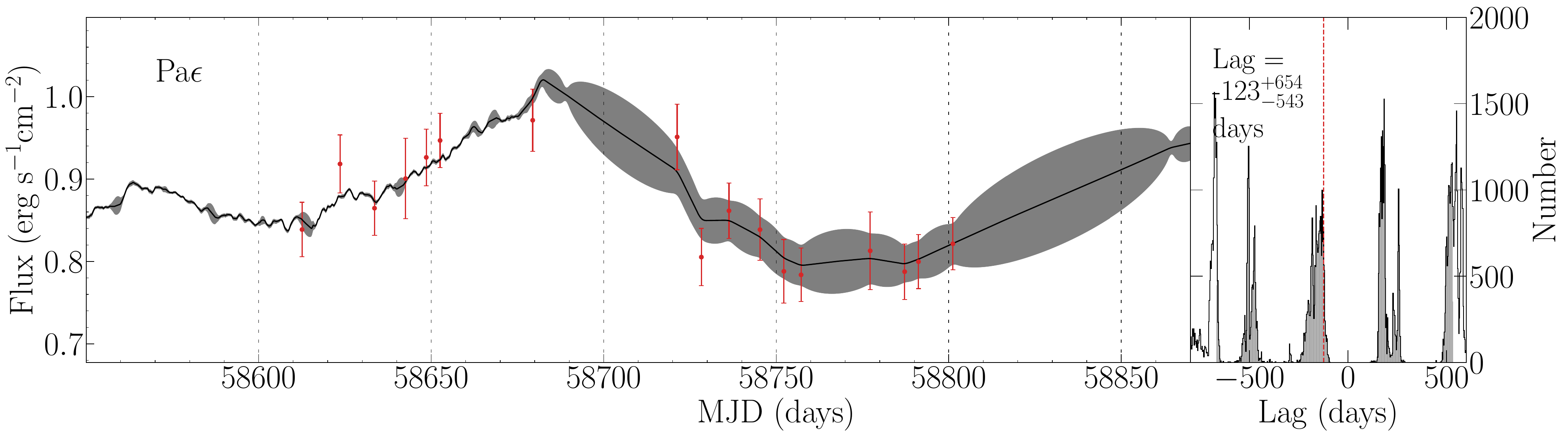}
}
\caption{\label{eps_jav_longlim} \javelin{} results for the Pa$\mathrm{\epsilon}$ broad emission line. Each run consists of $10^{5}$ iterations, with the determined lags displayed on the right hand of each panel. The 17 red data points represent enclosed emission line fluxes, and the continuous grey curve is the smoothed, shifted and scaled \javelin{} model of the `driving' $g$-band lightcurve displayed in Figure \ref{ztf_lco}. Lag limits placed at (-800 - +600) days.
}
\end{figure*}

\bsp
\label{lastpage}

\end{document}